# Adapting a Formal Model Theory to Applications in Augmented Personalized Medicine

Plamen L. Simeonov[1] and Andrée C. Ehresmann[2]

1) Charité - Universitätsmedizin, Berlin, DCGMS and JSRC, Germany;   2) LAMFA, Université de Picardie Jules Verne, Amiens, France.

**Abstract**: The goal of this paper is to advance an extensible theory of living systems using an approach to biomathematics and biocomputation that suitably addresses self-organized, self-referential and anticipatory systems with multi-temporal multi-agents. Our first step is to provide foundations for modelling of emergent and evolving dynamic multi-level organic complexes and their sustentative processes in artificial and natural life systems. Main applications are in life sciences, medicine, ecology and astrobiology, as well as robotics, industrial automation, man-machine interface and creative design. Since 2011 over 100 scientists from a number of disciplines have been exploring a substantial set of theoretical frameworks for a comprehensive theory of life known as Integral Biomathics. That effort identified the need for a robust core model of organisms as dynamic wholes, using advanced and adequately computable mathematics. The work described here for that core combines the advantages of a situation and context aware multivalent computational logic for active self-organizing networks, Wandering Logic Intelligence (WLI), and a multi-scale dynamic category theory, Memory Evolutive Systems (MES), hence WLIMES. This is presented to the modeller via a formal augmented reality language as a first step towards practical modelling and simulation of multi-level living systems. Initial work focuses on the design and implementation of this visual language and calculus (VLC) and its graphical user interface. The results will be integrated within the current methodology and practices of theoretical biology and (personalized) medicine to deepen and to enhance the holistic understanding of life.

## 1. Prelude

Human knowledge about the "nature of laws in nature" expanded in the past 50 years (Mikulecky, 2000, 2007; Maturana, 2000; Simeonov, 2010; Simeonov et al., 2012a). Researchers realize now that living systems exhibit essentially a non-linear dynamics featuring development at the edge of criticality, (Bak et al., 1987 ; Hankey, 2015). However, many structures and functions at the systems level overlap suggesting that they might not have clear physiological definition. Perhaps knowing the fact that *Biological First Principles* (Torday, 2017) define cellular freedom of action and cellular constraints which are the action central of biology/evolution might help developing more realistic models. But mastering the modeling of complex biological systems is still a serious challenge. Denis Noble has hosted prominent work on defining the problem (Noble, 2012, 2015). In this juncture he appealed for practical and sustainable models that unite the different levels of representation. To achieve this goal, researchers are asked to adopt a more holistic and 'organic' view of life that goes beyond general system theory (von Bertalanffy, 1950, 1968), systems biology (Kitano, 2001, 2002; Bard et al., 2013; Covert, 2015) and fuzzy set systems (Zadeh, 1965, 1975, 2004) into the realms of quantum interactions (Gurwitsch, 1922, 1944, 1947; Popp, 1992, 2003; Hameroff and Penrose, 1996; McFaden, 2002, 2011; Beloussov and Voelkov, 2007; Lozneanu and Sandoluviciu, 2008; Beloussov, 2008; Arndt et al., 2009; Levin, 2011, 2012)  and multiverse theory (Everett, 1956/1973, 1957). But along the way they face two special challenges. On the one hand, there are extremely complex information and energy flows in living systems. Characteristic for this approach is the appraisal of living systems balanced at multiple levels from molecules through cells, and tissues to organ(ism)s and (eco)systems. This assembly needs to be adequately comprehended and reconstructed in order to solve severe problems such as developmental and epigenetic disorders, autoimmune diseases, limiting and extirpating a virus outbreak, etc. On the other hand, the present specialized biomedical tools are certainly insufficient to bring down new theories to everyday practice easily and to investigate in silico. But also physicians and medical assistants, dealing with these issues lack a sufficient background in mathematics and computer science to first devise theories[1] and then model and test them with experiments in the way that physicists have been practicing this in their own domain for ages.

---

[1] Deriving a theory from ("big") data – induction – "is not a valid method for scientific proof" (Francis, 2017; Preface).

This is certainly insufficient to bring down new theories to everyday practice easily and investigate in silico and systematically highly abstract theoretical models and hypotheses. Instead, expensive, often ad hoc empirical studies keep dominating the research landscape, thus letting classical inductive system biological and bioinformatics methods be the single source for models derived from extracted fragments of knowledge out of averaged big data analyses that passed statistical thresholds. The situation becomes even more demanding when facing the challenges of personalized medicine. Therefore, two major goals need to be tackled on the way towards future research in biology and medicine:

i) bringing down relational mathematics such as category theory for practical use including non-mathematicians; this will not only create a powerful momentum for the development of mathematically grounded methodology in medical diagnostics, but also induce dramatic reduction of animal experiments in everyday clinical practice[2];

ii) demonstrating an elegant biomathematical solution to a relevant problem in complex systems biology and medicine that cannot be solved today, but lies at high confidence within the scope of current research: the generation and validation of an overarching and easily comprehensible realistic relational and shared model in personalized medicine, targeting a more reliable diagnosis and therapy accounting not only for the recognition and progression of specific diseases, such as cancerogenesis or a cardiovascular whole-system (heart, kidney, lungs, liver) tracing, but also for the entire personality of the patient.

The approach proposed here is theory-driven and *mathematical*, i.e. *deductive*[3]. Data is integrated a posteriori to validate the models. The ultimate goal pursued with this work is the development of a highly sophisticated decision support system for life sciences and medicine. In the following sections, we present and discuss the essential characteristics of such prototype architecture.

## 2. Outset

*Integral Biomathics* (IB) was proposed as a unifying framework for research in life sciences and medicine (Bateson, 2000; Simeonov, 2010; Simeonov et al., 2011; Simeonov et al., 2012a/b; Simeonov et al., 2013a/b; Simeonov et al., 2015; Simeonov and Cottam, 2015). This program collects the viewpoints of leading scientists, mathematicians and philosophers to approaches beyond the current state of the art, for devising a new paradigm for theoretical research in biomedical sciences towards defining a theory of life, (Edelman and Gally, 2001; Barberi, 2003; Pradeu and Carosella, 2006; Barberi, 2008a/b; Kauffman, 2009; Letelier et al., 2011; Pradeu and Vitanza, 2012; Longo et al., 2012; Kineman, 2012; Shiah, 2016; Soto et al., 2016; Mosio et al., 2016). Stepping on earlier research in cybernetics[4] and theoretical biology[5], this field has been developed since 2011 by over 100 scientists, from a number of disciplines who have been exploring a substantial set of theoretical frameworks. That work followed another significant EU funded project led by the applicant: the INtegral BIOmathics Support Action (INBIOSA, www.inbiosa.eu) which gathered experts, built collaborative infrastructure, identified key issues and triggered the promise of solutions (Simeonov et al., 2012a/b). Three large volumes constitute the collected current state of the art in the field (Simeonov et al., 2012a; Simeonov et al., 2013a; Simeonov et al. 2015). A fourth one is currently on the way, (Simeonov et al., 2017), for exploring the relations of Integral Biomathics to the Eastern thought traditions. These foundations are unprecedented. A coherent statement of the program for work emerged (Simeonov et al. 2012b; Simeonov et al., 2013b; Simeonov and Cottam, 2015). However, the competition among possible solutions and directions, though much clearer now, is not resolved. Multiple approaches are likely to be appropriate for different requirements and these will only be proven through test implementations. This paper discusses a specific technical realization of the Integral Biomathics program methodology, which is built on solid mathematical foundations developed in the past 30 yeas.

---

[2] Keeping in the same paradigm this model could also simulate the action of new medicaments on the target disease, which require clinical studies with human volunteers. In this way, these relational in silico models could avoid long time collateral effects, which are difficult to anticipate because of the variability of their onset. To the extent that dynamic categorical mappings integrate time, which is the subject of another related project proposal, this opens up an opportunity to test these assumptions by accelerating time.
[3] We take 'deductive' in the sense 'data-deductive'.
[4] the Macy conferences (1946-1953) with McCulloch, Pitts, Bateson, Shannon, Mead, von Foerster, Wiener, von Neumann and others.
[5] the Villa Serbelloni conferences (1965-1968) with Waddington, Thom, Goodwin, Conrad, Smith, Rosen, Kauffman and others.

IB is the continuation and extension of the research line traced by Rashevsky (1943a/b; 1944; 1954; 1960a/b), Waddington-Goodwin (Waddington, 1957; Goodwin, 1963; Waddington, 1968), Varela-Maturana-Uribe (1974), Rosen-Louie (Rosen 1958a/b, 1959, 1991, 1999; Louie 2009, 2013) and others (Simeonov, 2002c; Wepiwé, and Simeonov, 2006a; Kauffman, 1987, 2015). Its core insight is that the clue to understanding living systems is their development as 'organic' multi-level complexes, captured by means of *appropriate* biomathematical and biocomputational formalisms. This does not mean that using these theories should be an end in itself. The goal is to address truly eligible applications of mathematics and computation to biology. In other words, the quest is for those *patterns* in biology – called diagrams in mathematics – that can be informed by adequate mathematics and computation methods. At the same time, we can preserve the empirically observed relationships between the elements at many different levels all throughout these modelling transformations. However, it is not just a matter of making higher mathematics and theoretical computer science available for studying biology in a new way. It is a matter of finding just what (and which) kinds of fields fit in the problem descriptions and help solving them[6]. Examples of relations between mathematics/computation and biology that can be further developed within IB are:

- *Helical Geometry* – to investigate the DNA structure;
- *Topology* – to address the limitations of the Watson-Crick model; the discovery of the topoisomerase enzymes reflected in the differential geometry/topology of helices belongs to the modern understanding of DNA;
- *Knot Theory* – for studying recombination processes: tangle models;
- *Mathematics of self-reference and re-entry of distinction* (G. Spencer-Brown, N. Luhmann, L. H. Kauffman) – to study the replication of DNA, biological epistemology and other reflexive domains (Maturana and Varela);
- *Category Theory* – to understand whole system behaviour of organisms, *etc*.

In this sense, Integral Biomathics is substantially different from systems biology today (Wigner, 1960; Simeonov et al. 2012b; Bard et al., 2013; Hoffman, 2013; Salthe, 2013; Noble, 2015) and claims to be a new, extended branch of theoretical biology as it was envisioned some 50 years ago (Waddington, 1968). It comprises, inter alia, not only the *relational* aspect of theoretical biology (Rashevsky, 1943a/b; Rosen, 1958a/b, 1959, 1991, 1999; Louie, 2009; Kineman, 2012; Louie, 2013), but also its *experienced*, first-person, aspect (Brentano, 1874; James, 1890; Husserl, 1991; Heidegger, 1969; Merleau-Ponty, 1964, 1968, 2013; Simeonov et al., 2015; Ehresmann and Gomez-Ramirez, 2015) in its models. Published research in (Simeonov et al., 2015, 2013a, 2012a) provided the foundation for incorporating and implementing the requirements of (Simeonov et al, 2012b, 2013b; Simeonov and Cottam, 2015) and other key findings of the INBIOSA initiative. Previous efforts in the field coalesced on six key themes (Simeonov and Cottam, 2015) and two focused issues: i) the central importance of novel biology-driven mathematical and computational models of *dynamic multilevel complex systems* that sufficiently well match the "exo" and "endo" phenomena at hand, (Atmanspacher and Dalenoort, 1994): organisms, their conglomerations and internal processes of development and disease, and ii) the key role of *timing and synchronization* in biology characterized by the requirements for a new, integral notion of time for living systems (Matsuno and Salthe, 2002; Matsuno, 2012, 2015, 2016, Simeonov, 2015), comprising the phenomenological aspect of (Husserl, 1991; Heidegger, 1969; Waddington, 1957; Goodwin, 1963; Waddington, 1968; Varela, 1999; Vargas et al., 2013; Canales, 2015; Vrobel, 2015) and retrocausality (Pegg, 2008; Price, 2012; Matsuno, 2016). These now fairly discussed topics are currently guiding IB research. The first choice for advancing Integral Biomathics as an enhanced relational biology and "relational science", grounded on the ideas of Robert Rosen (1958a/b, 1959, 1991, 1999) and his followers Louie (2009, 2013), as well as Ehresmann and Vanbremeersch (1987, 2007), became Category Theory[7] in its special instantiation for modeling multilayered living systems. The second choice is the mathematics of recursion and self-reference (Kauffman, 1987, 2015).

---

[6] We should keep in mind that there are also incomputable branches of physics and biology, e.g. (Longo, 2012).
[7] Let us note that, in the past, CT has been criticized for its limited capability to model emergent and quantum phenomena in living systems using organizational closures (causal entailments) and modelling relations known from the works of Rashevsky (1943a/b; 1944; 1954; 1960a/b) and Rosen (Rosen, 1958a/b, 1959, 1991, 1999) on Relational Biology and (M,R)-systems. Yet recently, CT has made significant progress with new extensions and syntheses suggested, e.g. in the research of Letelier et al. (2011), Kineman (2012), Louie (2009, 2013), Soto et. al.(2016) and Mosio et al. (2016), and in particular in the work of A. C. Ehresmann and J.-P. Vanbremeersch (1987, 2007) later referred in this paper.

In this context, Simeonov and Ehresmann have proposed a formal biomathematical and biocomputational research framework: the Wandering Logic Intelligence Memory Evolutive Systems, *WLIMES*, (Ehresmann and Simeonov, 2012). It is based on the complementary synergy between i) a *non-axiomatic* (i.e. "rigid but flexible", (Wang, 2006) situation and context aware spatiotemporal logic, *WLI* (Simeonov, 2002a/b), and ii) a dynamic category theory[8] for multi-level, multi-agents, *MES* (Ehresmann and Vanbremeersch, 1987, 2007, 2009a/b). It represents a sophisticated hybrid design methodology rooted in mathematics and computation for modeling conceptualization in multi-layered complex living systems. The WLIMES perspective on Integral Biomathics was presented in two of the preceding publications (Ehresmann and Simeonov, 2012; Simeonov and Ehresmann, 2017). The latest contribution (Simeonov and Ehresmann, 2017) carefully examined the main principles and characteristics of, successively, MES, WLI and WLIMES in relation to basic philosophical concepts and archetypes[9] from Eastern thought traditions. The present article is another incremental step refining the exemplified Integral Biomathics approach in the face of WLIMES towards its practical implementation in the biosciences and medicine.

## 3. Theoretical Framework

This section presents the theoretical foundations on which we are constructing formal models of complex biological systems and devise methodologies for their investigation.

### 3.1. The Wandering Logic Intelligence (WLI)

The Wandering Logic Intelligence (WLI) is a formal computational theory, which was used for designing active self-organizing mobile networks based on ad hoc situation and context based knowledge acquisition and examination. It is based on temporal logic of actions, s. (Lamport, 1994; Simeonov, 2002c; Wepiwé, and Simeonov, 2006a; Simeonov, 1999; Wepiwé and Simeonov, 2006b). An essential characteristic of this model is its inherent ability to instantly spread out data and instructions about structural changes among the individual elements of a communication network by encoding 'genetic' instructions for migrating and replicating these structures as virtual computing environments. This is achieved via active information packets – as network genes, N-genes (Simeonov, 2002a, 2002c), (Fig. 1). The latter represent the functional analogy with a living system while being primarily focused on implementing the concepts of virtual communication infrastructures, information encoding and context-based exchange between the participants.

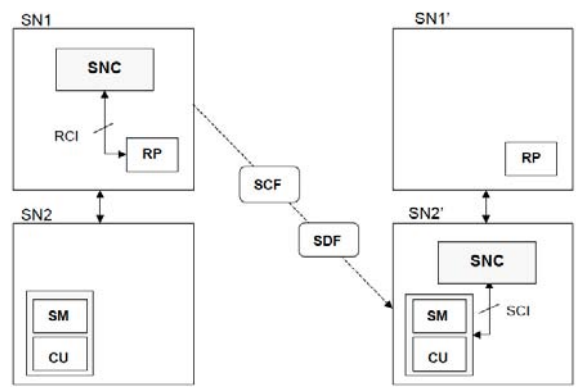

**Fig. 1:** *The migration of functions (wandering logic) between network elements (ships) implemented as intelligent demand-oriented utilization of resources through the exchange of viral active packets (shuttles) to realize the evolution of virtual living infrastructures, (Simeonov, 1999, 2002b).*
<u>Legend</u>: SN – Service Node, SNC – Service Node Controller, RP – Resource Platform, RCI – Resource Control Interface, SCI – Service Control Interface, SM – Switch Matrix, CU – Control Unit, SCF – Service Control Function, SDF – Service Data Function; left-hand side of the image – configuration of the architecture wandering network architecture at the moment t before the "wandering" of the functions SCF and SDF from SN1 to SN2 at the moment t' (right hand side of the image).

---

[8] The *holobionts* (Margilis, 1991; Mindell, 1992; Torday, 2017) which self-organize to live and co-develop in symbiosis could be one particular case of this organizaton, but they do not cover the social and cognitive systems, which are the main application of MES.
[9] an inherited idea or mode of thought in the psychology of C. G. Jung that is derived from the experience of the earliest human ancestors and supposed to be present in the collective unconscious of the individual [Oxford Living Dictionary & Merriam Webster].

Thus WLI represents an evolving network architecture which is composed of dynamically reconfigurable network elements (called "netbots" or "ships") generating and exchanging information about themselves and their surrounding environment (close neighborhood or "local landscape") by means of active information packets (called "shuttles") containing data and executable code to process them in the form of n-genes. In a Wandering Network (WN) shuttles are used for signaling and transporting various kinds of information (physical, algorithmic, topological, etc.), cf. Appendix A for realization details. Initially, temporality in WLI has been modeled as external linear clock time.

Netbots in WLI do not exhibit an explicit hierarchy. They are heterarchical (McCulloch, 1945; von Goldamer et al., 2003). Their landscapes at a *given instant of time*[10] demonstrate a temporarily available internal composite hierarchy of structures and functions of their building elements interlinked with other elements and groups of them, including elements inside them, in particular elements inside remote netbots. This composite hierarchy is changed stepwise by that outcome of processing the shuttle information in combination with other internal and external exchange within the individual netbots and other components of the system.

In summary, a WLI realization implies a multiplicity of evolving network gates, which can be regarded as an abstraction of an ecology system of living organisms maintaining its equilibrium. By definition the WLI nodes/gates cooperate to realize self-stabilizing network architectures, (cf. Appendix A). However, they can also compete for a resource/function by involving some special reservation policies based on monitoring their shuttle traffic. More details about these and other characteristics of the Wandering Logic Intelligence are given in (Simeonov, 2002a/b). The WLI implementations in mobile multimedia ad hoc communications were presented in (Simeonov, 1999; Simeonov, 2002c; Wepiwé and Simeonov, 2006a/b).

### 3.2. Category Theory (CT)

Category Theory (CT) (Eilenberg and MacLane, 1945; MacLane, 1998; Johnstone, 2002; Awodey, 2006; Borceux, 2008) is a formal domain of mathematics at the frontier between mathematics, logic and meta-mathematics. It provides: i) a language which tries to uncover and unify the operations made in different mathematical branches (such as abstract algebra, topology, geometry, etc.); (ii) a general notion of mathematical structure (e.g. category of sets/groups/graphs/fields/rings etc.); and iii) study of "universal properties" with the help of specific tools (such as adjoint functors, colimits, etc). A *category* is a directed *graph* with a supplementary structure, namely an internal law to compose successive arrows of the graph; a vertex of the graph is called an object and an arrow is called a morphism. For instance, in *Set* (the category of sets), the objects are (small) sets and the morphisms are maps between them. However, there are also 'small' categories such as monoids (categories with a unique object) or categories associated with an ordered set (categories with at most one morphism between two objects), thus giving much more representational freedom.

By construction Category Theory offers a 'relational mathematics' that is essential in biology (Rosen, 1958a/b, 1959, 1991, 1999; Louie, 2009; Kineman, 2012; Louie, 2013). It emphasizes the relations between objects rather than the objects themselves. Indeed an object of a category is characterized not by its 'ontological structure' but by the set of morphisms arriving to it (Yoneda Lemma, (MacLane, 1996)), which corresponds to the different operations (functions) in which it participates. In this way, CT represents a useful tool to define highly abstract relational models of complex systems[11].

In particular, Category Theory helps in solving the following predicaments:

---

[10] Defining this specific context by means of an explicit second order multidimensional logic is a key challenge in biology.
[11] While many of the works using category theory are qualitative, Haruna and Gunji (2009a/b) succeeded in explaining some traits of biological networks quantitatively.

(i) The *Binding Problem*: how to bind simple interacting objects together to build a complex object, which is "a whole that is something else than the sum of its parts"? The categorical notion of a *colimit*[12] – Kan 'inductive limit', (Kan, 1958) – provides the answer.

(ii) The *Emergence Problem*: how to form objects and processes of increasing complexity satisfying emergent properties? An answer to this question is given by the "Complexification process", (Ehresmann and Vanbremeersch, 1987, 2007).

Detailed treatises about CT can be found in (Eilenberg and MacLane, 1945; C. Ehresmann, 1966; Bell, 1981; Landry, 1995; Awodey, 2006; Borceux,, 2008; Schmidt, 2011; Schmidt and Winter, 2014, Riehl, 2016). Two category-theoretical approaches are of particular interest for the formal implementation of Integral Biomathics: Memory Evolutive Systems, MES (Ehresmann and Vanbremeersch, 1987, 2007), and Relational Biology, RB (Louie, 2009, 2013), with the last one being a continuation of Robert Rosen's work on the subject (Rosen, 1958a/b, 1959, 1991, 1999) and on anticipatory systems (Rosen, 1985; Louie 2010; Ehresmann, 2017). Both of them represent specific incarnations of CT for biology as synergetic syntheses of a number of mathematical theories (partially ordered sets, lattices, graphs, categories, adjacency matrices, interacting entailment networks, etc.).

### 3.3. Memory Evolutive Systems (MES)

MES on their part introduce new categorical notions: (i) *Hierarchical Categories* to represent a multi-level organization (from cells up to tissues, organs and systems in the living); (ii) *Evolutive Systems* (as a family of categories indexed by Time and connected by partial functors measuring the changes between them) to account for the evolution of multi-level complex systems such as biological systems; (iii) The *Multiplicity Principle* (generalizing the degeneracy property of biological systems) at the basis of emergence; (iv) The *Complexification Process* describing the emergence of higher order components connected by 'complex links' which represent emergent properties.

The configuration $S_t$ of the system S at moment t is represented by a hierarchical category where objects correspond to components at the moment t, and links to channels for their interactions, (Fig. 2). The objects are divided into levels so that C of level n+1 has an internal organization into a pattern P of linked components of lower levels, which it 'binds' so that C and P have the same functional role. C is modeled by the colimit of P. Thus, an object C of level *n*+1 in a hierarchical category **H** is the colimit of at least one pattern P of interconnected objects of lower levels. It is called a 'holon', (Koestler, 1967, 1978), representing alternatively i) a part of, or ii) the whole itself with respect to the objects of its lower level decompositions. The object C is 'simple' with respect to an object of level n+2 if it belongs to a decomposition of this object; C admits also at least one ramification[13] down to level 0. The *order of complexity* of C is the shortest length of its ramifications.

A MES is an evolutionary system in which the change from $S_t$ to a later configuration $S_{t'}$ is represented by a 'transition' functor from a subcategory of $S_t$ to $S_{t''}$ so that the whole system is categorically modeled by a Hierarchical Evolutive System (introduced by (Ehresmann and Vanbremeersch, 1987)). An interesting aspect of MES is that in addition to its representation of the system and its changes over time, it also describes how this dynamic is internally generated by a network of internal agents, called *co-regulators* (CRs), which have "different points of view" upon the system organization and its operations, i.e. different 'landscapes' reflecting their own situation inside the system: they form a heterarchical organization, (McCulloch, 1945), where a CR of level n does not aggregate lower level co-regulators though it might send commands to them. The CRs operate with the help of the Memory of the system (cf. Appendix B), which they develop over time, allowing for increasing knowledge and adaptation to external changes.

---

[12] For instance, let us take a category C whose objects are molecules and atoms, and morphisms modelling chemical interactions between them. We can look at a molecule as the set of its atoms and their chemical bonds. However, the morphisms between molecules do not correspond to maps between these sets, but represent emerging properties at the molecular level depending on quantum level properties. This is categorically represented by the fact that a molecule becomes the colimit in C of the pattern consisting of its atoms and their chemical bonds.

[13] A *ramification* of C consists of a decomposition P of C, then of a decomposition of each Pi of P, and so on down to patterns of level 0.

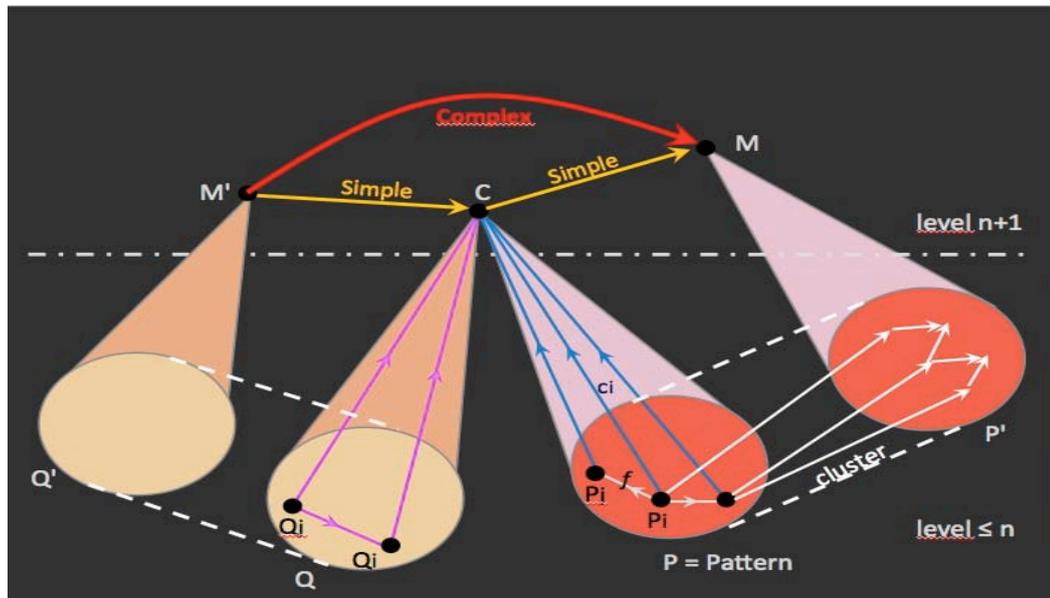

**Fig. 2**: *The Construction of the next level of a hierarchical category via simple[14] and complex[15] links (adopted from (Ehresmann, A. C., Vanbremeersch, J.-P. 2007)).*

A co-regulator represents an evolutive subsystem of the MES with its own function, complexity and rhythm. It operates as a hybrid system, (Branicky, 2005), with its own discrete time-scale delimiting its successive steps on the continuous time-scale of the system. At each step, the information received by a CR via the active links arriving to it during the step is processed in the *landscape* of CR at the moment t, which is modeled by an Evolutive System having those links for components. Using the memory, an adapted procedure Pr is selected on the landscape, and the corresponding commands are sent to effectors; that starts a dynamical process whose result will be evaluated at the beginning of the next step; if the objectives are not attained, there is a *fracture* for CR. A main cause of fractures is the non-respect of the *synchronicity laws,* which a CR must respect (they relate the length of a step with the propagation delays and stability spans of the components of the landscape; cf. (Ehresmann, and Vanbremeersch, 2007, chapter 7).

To account for the types of changes between configurations of the system, such as those taking place during morphogenesis, the transitions between configurations are modeled by iterated *complexifications* in MES which (thanks to the Multiplicity Principle) may introduce more complex new components connected by complex links. Such a complexification consists of operating changes of the type: addition, suppression and/or combination of patterns of interacting components. This procedure gives an interesting categorical approach to represent an 'organic' universe compatible with Kauffman's idea of an "open universe" (Kauffman, 2009). It is different from the construction given by Graves and Blaine with the Algos approach (Graves and Blaine, 1985a/b; Graves, 2013) where Algos designs a large topos (in the sense of Lawvere-Tierney (Lawvere, 1971)). In fact their approach would correspond to construct complexifications internally to the topos Algos instead of constructing 'free' complexifications (i.e. solutions of a universal problem), which allow for the 'openness' of the universe.

A detailed discussion of CT and MES characteristics is given in (Ehresmann and Vanbremeersch 1987, 2007; Ehresmann and Simeonov, 2012; Simeonov and Ehresmann, 2017). Some of their applications in life sciences are presented e.g. in (Ehresmann and Vanbremeersch, 2011; Ehresmann, 2017a/b/c), cf. also Appendix B.

---

[14] A (Q', Q)-simple link g from M' to C transmits only information already mediated through the components of M and C in Q' and Q, since it binds a collective link from Q' to C, of which all the individual links are mediated by Q.

[15] *A complex link is obtained by composing simple links binding non-adjacent clusters. The patterns P and Q in Fig. 1 are homologous patterns having the same colimit C. The link g from M' to C is a (Q', Q)-simple link binding the cluster G, and the link g' from C to M is a (P, P')-simple link binding the cluster G'. If Q and P are homologous but not connected, the composite link gg' is generally not a (Q', P')-simple link, because it does not bind any cluster from Q' to P'.*

## 3.4. WLIMES as synergetic synthesis for biology-driven mathematics and computation

WLI was extended in 2012 by Ehresmann's dynamic category theory, Memory Evolutive Systems (MES), to become amenable for modeling living systems (WLIMES, (Ehresmann and Simeonov, 2012; Simeonov and Ehresmann, 2017)). Co-regulators (CRs) in MES correspond to specialized subsystems of elements, which are not necessarily disjoint, i.e. an element can belong to multiple CRs. Thus, CRs correspond to different (virtual) levels of structural and/or functional organization of the netbots. This organization is defined by the internal computation processes inside the netbot and by their external information delivered through the arriving shuttles. In this way, WLI determines the operational[16] semantics of the system model, whereas MES – its denotational[17] semantics. While the components form an explicit hierarchy in MES, the CRs don't form a real hierarchy; their level comes from the level of their components, but a CR of level *n* does not aggregate CRs of lower levels.

Herewith, the six WLI principles (*Dualistic Congruence, Self-Reference, Multidimensional Feedback, Pulsating Metamorphosis, Resource Availability and Utilization, Second-Order Logic*[18]) define the overall development of the network infrastructure, or the living organ(ism) in a biomedical context, whereas the transitions between netbots and their constituents execute a double function. On the one hand they illustrate such operations as addition, loss and binding of functional components inside a WLI closure or (sub)network. On the other hand, they represent unidirectional or bidirectional channels along which shuttles are transported. In MES, the links correspond to directed channels between the different components, which can be inter-levels (up or down) or intra-levels. The latter were implied in WLI in connection with the transmission of different types of shuttles. One of the differences is that in MES there are also components (for instance in the memory, but not only there), which don't belong to CRs. The operations of a CR, such as the formation of its landscape, must also account for these other components. This is not the case with WLI yet and needs to be considered. The Wandering Network suggests a dynamic hierarchy within its multiple closures/(sub)networks.

The netbots are cooperative; they negotiate their interplay/communication. They could be regarded as CRs in an emerging/development stage. Once the established channels between functional components/netbots become more frequently used, they can build semi-stable and permanent CRs and higher-level structures of them. Hence, the CRs of a MES correspond to netbots or (virtual) clusters of them operating as units and always participating a WLI (sub)network. However the CRs can be competitive and even conflicting, and not always cooperative. The main problem in MES is to resolve the 'interplay' among the CRs that, at a given time, must harmonize their local dynamics into the global dynamic of the system. In some regular enough specific cases, this interplay can be solved using results on hybrid differential systems (Branicky, 2005), but a general solution is all the more difficult since the Multiplicity Principle gives many degrees of freedom to this interplay.

This might be resolved using a trade-off scheme that reshapes CRs by introducing 'soft' and 'hard' types of netbots with varying threshold levels that can be adjusted via internal/external signaling through the shuttles for the good of the whole system. In this way, a sophisticated "rigid flexibility" logic (Wang, 2006) will be applied from the WLI operational semantics side with both kinds of behavior made possible at the extreme ends. This can be adjusted for each CR/netbot individually and temporary. An example of the situation and context aware wandering network is given in Appendix A.

---

[16] "In an operational semantics we are concerned with *how* to execute programs and not merely with what the results of the execution are." (Nielson and Nielson, 2007, Ch. 2). „A computational model is a formal model whose primary semantics is operational; that is, the model prescribes a sequence of steps or instructions that can be executed by an abstract machine, which can be implemented on a real computer." (Fisher and Henzinger, 2007).

[17] A mathematical model is a formal model whose primary semantics is denotational; that is, the model describes by equations or graphical entities (e.g. nodes and arrows) a relationship between quantities (sets) and how they change over time. The equations (or the graphical entities) do not determine an algorithm for solving them, (Fisher and Henzinger, 2007).

[18] The last, sixth principle was added recently to capture newer developments (Abramsky and Coecke, 2004; Abramsky and Duncan, 2004; Jerzak, 2009; Courcelle and Enmgelfriet, 2012; Goranson and Cardier, 2013; Goranson et al., 2015; Simeonov, 2015; Matsuno, 2015; Cardier et al., 2017). It is related to the logic at work inside the system, i.e. between the particular hierarchical levels, which does not necessarily obey a binary law (e.g. multivalued and intuitionistic logic). Instead, there is a *second-order logic,* which successively turns the preceding predicate of an expression at a level of hierarchy into the succeeding subject at the next level expression, thus allowing for additional quantifications of the individual levels. In this way we observe a sort of bottom-up gradient in the flexibility of the logic, multi-valence, (Boole, 1854; Piantadosi et al., 2010; Cabrer et al., 2014) from the elementary physicochemical level to the behavioral level of an organism.

A WLI cluster/node can cease to operate/exist as a result of the interpretation/processing the information contained in incoming shuttles. In this way a degradation, 'aging' and even death of some parts of the wandering network is possible, thus enabling its replacement with new functional structures and links/channels between them. In the same manner, the regeneration of same former dead areas of the network occurs, once they involve living components, i.e. those at the edges of the network which maintain at least one connection to another netbot and capable to process its shuttles. The realization of such repair mechanisms depends on the particular 'regeneration' policies; the latter are matter of future investigation. This recomposition is also possible in a MES. Both components and CRs may disappear either completely or through replacement by others, and new ones can be created (by aggregation of patterns) through the complexification process. Also, the memory in MES is centralized, with each CR having its own differential access to it, whereas in WLI it is distributed among the netbots, their components and the exchanged shuttles.

With respect to the (individual) duration of these processes, in MES there is a kind of duality between the situation at the beginning moment $t$ and at the end moment $t'$ of a step of a CR, caused by the information gathered in its landscape during the step from the *point in time*[19] $t$ to $t'$ (through incoming 'shuttles'), and the selection of a procedure to respond by the CR. On the other hand, the netbot also causes changes in the shuttle while processing it, thus influencing changes in other netbots receiving that shuttle at a later moment[20]. This can be regarded as another incarnation of the dualistic congruence principle. In MES the commands of the procedure selected by a CR transmit information (through 'shuttles') to other CRs, for instance to effector CRs such as the effectors of a 'muscular' command.

In summary, a WLI realization implies a multiplicity of evolving gates. By definition, the WLI nodes/gates cooperate to enable a self-stabilizing network architecture. However, they can also compete for some resource/function by involving some special reservation policies that can be transmitted by means of shuttles. The intelligence/plasticity of the system at the moment $t$ is the instant result/response of the interplay between its constituting elements (components, netbots and interworking clusters of them). In MES the interplay is realized through the procedures of the different CRs at a given time, and its flexibility comes from the multiplicity principle. However, the different temporalities of the CRs require careful planning about how the signaling and synchronization mechanisms are going to be realized. A detailed discussion of the WLIMES characteristics can be found in (Ehresmann and Simeonov, 2012; Simeonov and Ehresmann, 2017).

**4. A case study: shared whole slide image analysis and diagnosis support in oncology**

**4.1. The integration of multiple views into a general framework for personalized medicine**

We distinguish between different stages of evaluation of a pathology/patient, which eventually develop into a framework to systematize rare (divergent) versus general (normal) states of the investigated case:

- the treating/attending physician;
- peers/colleagues in the same field (for instance, physicians in clinics);
- specialized peers/colleagues in complementary fields (e.g. pathologists, radiologists, etc.);
- the patient (incl. his/her personal 'feeling' of the situation) and his/her family uncovering subtle and hidden relations), i.e. phenomenological aspect;
- other patients with the same or similar disease (uncovering also hidden/subtle relations).

---

[19] This is another temporality term that needs to be carefully defined under the novel concept of integrative multidimensional time.
[20] Again, simultaneity or sequence in time must be defined differently within the new concept of integrative biotime. A 'global' linear time is in view of the phenomenological consequences not the appropriate option.

The entire review procedure has to be organized like a multiple nested spheres system with the same center (source of data, patient case). All parties have different approaches to the situational 'design', they make different observations and focus on different things to contribute to creating complementary aspects of the whole system/situation[21]. There are also diverse relations between cultural background, personality and identity (in terms of being part of a group), which also play a role in the creative process of deriving a shared diagnosis/hypothesis. Sometimes these generalizations are not beneficial, in the case of stereotypes that make it difficult for a physician to accurately assess a person's individual malady, but most often these make it possible to consider biological or lifestyle constraints that might inform on a disorder. Each single viewpoint has to be parametrized, adjusted and normalized with the others to allow a joint case construction.

WLIMES could provide a system support for realizing such a construction and analyzing the results versus the individual stages / target groups as follows:

1. Success for the individual physician means to have a compositional diagnosis (hypothesis) designed with the help of the WLIMES system that is coherent with one's own personal assessment.
2. Success within the first group of peer colleagues from the same field means to have a *good creative process* referring (in a "constructive or destructive way") to the discipline discourse and its context to deliver a diagnosis/hypothesis that is consistent with the approach of the individual physician.
3. Success with the second group of colleagues from related fields means to have developed a universal *artistic* language and an aesthetics that is not depending on the knowledge of the process and its context.

The last three groups (individual patient, groups of patients with the same or related diseases, non-experts) means that the diagnosis/hypothesis is derived in a way that it contains a transparent message that is easily understandable, recognizable and identifiable. These groups are usually excluded in medical analyses, but in general ordinary humans can also follow the detailed explanations of the physician (if s/he does not speak lingo). They cannot always see the reasons, but they understand the arguments (the logic) when watching the images led by their own reasoning system. All different groups (and individuals within the single groups) make different observations and focus on different aspects of the case to contribute complementary details to the whole system/situation. There are also diverse relations between cultural background, personality and identity (in terms of being part of a group), which also play a role in the creative process of deriving a shared diagnosis/hypothesis. An 'augmented' hypothesis/diagnosis by multiple agents is generally good when it reaches a sort of "compromise" between the individual judgments.

### 4.2. Medical image interpretation through the eyes of the expert

*The ultimate goal* of the WLIMES approach is the development of a new kind of intelligent imaging decision support system for personalized medicine based on the integration of mathematics and computation for model building and hypotheses test. This section proposes to develop a high-end graphical deduction system supporting human-computer interaction and capable of reproducing sufficiently well and accurately the landscape of a pathologist who tries to interpret the images in a Whole Slide Image (WSI) in terms of undergoing processes. The (artificial) reasoning process along with its ontologies can be represented as a hierarchical category capturing a number of levels:

---

[21] Personalized medicine based on DNA analyses has an essential difficulty when applying the above method. On one hand it requires averaged pattern of a DNA sequence found in a given disorder. But in practice, the deviation or individuality is ignored and removed. Then, once the trend for a given disorder is determined as an average, the deviation is regarded as essential personality. However, the information of an averaged sequence of DNA is different from the average values of e.g. height of bodies. Embedding the various perspectives of the experts into one shared perspective on one individual as reported in the WSIs example in Section 4.2 requires the integration of multiple expert views upon the entire molecular pathways network of this particular individual to derive an individual diagnosis which means an increase of complexity. Therefore, the molecular pathways should be organized as far as possible in terms of hierarchical structures according to the MES Multiplicity Principle.

molecules, proteins, sub-cellular elements, cells, organized cell-clusters (glands, muscles, nerves, stroma, etc.), organ. This is more or less a standard procedure. The remaining tough question, however, is how to characterize the morphisms/links between these levels expressed in terms of subtle deviations from the normal forms. To be useful for the diagnosis, to both objects and links should be attached (sometimes 'fuzzy') geometric properties such as, symmetry, sharpness, convex/concave form, gradient/slope of lines, etc. and its relative closeness to other glands (related to the size of its diameter or 'skeleton' middle line distance from the boundary). Also the measured density of objects in a specific region is essential to make certain distinctions, e.g. a threshold of glands vs. 'empty space' (stroma) of 50% to justify if two glands are 'near enough', or the one is 'larger than usual'. Thus categorical tools can organize and automate the reasoning process, but they cannot help extracting specific properties from the image in the way the human brain realizes this. This suggests the necessity for integrating a) a low-level feature-based quantitative analysis that identifies, structures and differentiates the image in terms of "primitives" (nuclei, cells, glands, vessels, etc.), and b) a high-level qualitative (phenomenological) analysis and fuzzy/vague (intuitive "impressionistic") weighting of the image after filtering out the noise, this level corresponding to describe the landscape of the expert-pathologist.

*The reasoning process* of such an expert is very different from the one of a computer scientist or a mathematician. It realizes some sort of tradeoff between quantitative and qualitative modelling, which follows the patterns of an architect, a painter or a musician, involving the comprehension of aesthetic features and deviations from them. The expert essentially tries to guess and reconstruct the processes behind a "frozen" fossil stamp/record or "weather map" of the cellular/tissue configuration in the histology probe. The way in which pathologists describe the structures they observe during a diagnosis is remarkable, even artistic. They use e.g. metaphors and associations to describe the biological processes behind the observed forms. But such an analysis uses only two or three of the extracted features to be realized automatically, since the consideration of all of them in the way an expert considers them would take a lot of effort to disentangle the principles on which the logic of such an observation of one individual expert is built on (incl. features such as mitoses in different regions of the stroma and glands, and stages of development). We thus also refer to theories of conceptual change and practices in the arts to learn how information is carefully vetted for relevance and left out.

Sometimes, it is difficult to exactly define the position, size and form of the individual formations such as cells and glands, because of the uncertainty of the object recognition. For instance, the boundaries of objects may appear fuzzy and blurred. It can be difficult to distinguish between convex or concave lines in a local devolution. Objects such as glands are tending to merge in the case of tumors. One cannot distinguish by a simple algorithm how many of them are clustered and how strong/dense is their overlapping to judge about the biological processes and physical tensions in the tissue. In addition, there are diverse strange forms that need to be distinguished and classified according to sets of basic features.

The question to be answered is: *how to characterize the relations, i.e. the morphisms/links between these irregular objects/structures*? We have to partially invent these characteristics and define the criteria for their determination following the intuition and the virtues of the expert. Usually, a pathologist is analyzing a whole slide image in a top-down manner by zooming into more detail where s/he discovers a couple of interesting characteristics that can be set in a relationship about the hidden nature of the phenomenon. The feature selection and their weighting depend on the level of resolution and on the homogeneity contrast with the surrounding environment. For instance, some structures may percolate in an unusual manner into others. Since images are taken in the past, it cannot be said much about their pertinence at the present moment unless some conventions about the speed, acceleration and direction of tumor growth are met based on the morphological characteristics, which represent underlying processes. An analysis may include the replacement of the temporal dimension with spatial ones in multi-level systems, if appropriate, e.g. when tracing the development of a skin tumor. Of course, the individual diagnoses contributing to the constructed and shared diagnosis should be weighted in some reliable manner, e.g. through validation with real data to obtain a robust model. The individual judgements could be sorted out e.g. by using some ranking algorithm.

The interesting issue about using the WLIMES formalism, (Ehresmann and Simeonov, 2012; Simeonov and Ehresmann, 2017), to structure and assess the composition of a Whole Slide Image (WSI) from biopsies is that it can mimic the reasoning of the human brain in *linking multiple (sometimes disjoint) features about the size, form, density*, special resolution, etc. of such clusters of objects (nuclei, cells) *to match the dynamics of the underlying processes* and deliver a set of hypotheses that can be validated against each other, on the way to deduce a case diagnosis.

The first tier of characterization can be defined as follows:

**feature** := situational context (qualitative, quantitative) of 2D image/object layout (
        ( feature_1 → supposed_underlying_process(es)_1);
        …
        (feature_i → supposed_underlying_process(es)_i);
        (feature_n → supposed_underlying_process(es)_n)
    )

The second tier of characterization is organized as follows:

**composition** (features_1_to_n)     → supposed complex of underlying processes
        → inference of a thread of reasoning based on the combined effect of features

What follows next is the design of multiple compositions, i.e. inference threads (one for each individual case and agent), their combination and weighting to build an individual diagnosis.

### 4.3. Deploying WLIMES to derive an interactive diagnosis or hypothesis

We wish not to build an expert system that replaces a pathologist, but one that sorts out the images (thousands and millions of them) with suspicious regions of carcinoma to be later diagnosed by a human expert. The task here is to model the thinking of the expert when studying and annotating a WSI, e.g. through diagnoses of recorded video clips, texts or narratives.

The first step towards designing such a front-end automatic image analysis system is to find a way to link images/words in descriptive ontologies, organised within a hypergraph or dynamic category built during a sample trace of an expert and derive the basic reasoning rules from it. This can be done several times for multiple images of the same phenomenon evaluated by multiple experts to obtain a set of local landscapes/opinions as decision chain. Then, once a system becomes sophisticated enough, e.g. after it has gone through a deep learning session based on such decision chains, if one takes out descriptions or images the system could "propose/guess" what a description could be or what an image could be. Such an analysis system would read an image and pick out evidence of anomalous and normal processes based on the shapes.

Our goal is to make a series of traceable judgments through multiple interactive exchange levels between the diagnostics support system and agent groups that can be addressed to provide a logically shared compositional evaluation of the individual case. The stages of pathological image analysis usually comprise:

- individual pathologist,
- other pathologists in the same field,
- peers / colleagues in related fields (e.g. radiology).

A WLIMES description of the realization of the first two stages (individual physician and peers/colleagues) is discussed in Appendix B. The third stage requires the case interpretation from a different viewpoint, often involving different kinds of image data, but the principle of merging views is the same.

In summary, we do not need a single recipe/algorithm but a series and combinations of algorithms that access the same data source. There is a "weighted deep learning" process involved in this design process: the more the images are jointly "judged" by those series of algorithms matching the individual agent groups, the more are they related to the solution of one's individual case/problem/context. Reasoning and hypothesis/diagnosis design algorithms should be developed to match (different) categories/subjects/agents and be specific but capable of accessing the same source data folder of scanned images and to communicate (exchange) information about situation analyses made in a such way that the folder (image) is readable (open) to all algorithms and subsequently augmented by them. This is the overall concept of the Charité Digital Pathology Platform (C-DPP).

**4.4. Causal reasoning from the point of view of narratives about image data**

The construction of hypothesis scenarios during a diagnosis process can be compared to the construction of narratives and characterized by:

- a tier above all descriptive documentaries based on principles of visual form and design that manages the tiers beneath it;
- a framework of rare (divergent) versus general (normal) states observed in an image.

In a narrative, what is expected/anticipated to see in normal states helps identifying the most important anomalies; this is the *causal agent* (CA). There can be a number of causal agents in a machine diagnosis scenario. They can interact with each other if shared data are used. If we adopt the convention that a pathology case is characterized and ranked by features that shift a situation away from the 'normal', the scenario for analyses is organized according to the following scheme:

1. The way of extracting essential characteristics from visual images (and of course, linguistic too, which can involve phenomena beyond our perception) is innate for an individual's specialization and imagination. To derive the simple "message" from a complex context is a scout's skill. This is done iteratively by shifting the attention in different areas of the image, not in a chaotic manner, but in an organized one, also individual.
2. The process is analogous to movement between 'stepping stones' of expected and anomalous elements (put a context) to understand their relationship, and ultimately identify what is happening in the overall situation. It imitates eyes shifting around a Whole Slide Image in which it is difficult to tell whether there is an anomaly[22] (e.g. a lesion, a displaced vessel, a cancer, etc.) or not. After looking at a *combination of factors* together, and matching them with known processes in the human mind, a characterization/judgement is made (illness or not).
3. Crucial to analysis and deduction are the **multiple threads of feature extraction** and reasoning which one follows in parallel with time progressing. Jumping from one pair feature-cause (underlying process) to another is supported by short-term memory always present and accessible to all threads to support the decision-making process, the evaluation and the ranking of the individual paths.
4. What happens in the tier above the 'stepping stones' is that the anomalous and the general are bound into one artefact – Goranson et al. (2015) refer to this as united artefact as a situation, based on their approach which uses Situation Theory. This level manages the *traversal of the 'stepping stones'* of discrete information, as it appears in the river of temporal flow, enabling us to cross from the one bank to the other (metaphorically expressed). This is an 'aesthetic tier' of symbolic characterization, which is a core element of painting, architecture, music, etc. and guided by such visual design principles as harmony, symmetry, smoothness of curvature, etc. which let us distinguish between the nature of different forms, i.e. not just normal and abnormal, but also forms that suggest change, or that a change is coming, or that decay of some kind is underway. The artefact as a 'design' (Cardier et al., 2017) is unique enough to reflect an expert's own originality, yet expressed in a way that is widely accessible to others.

---

[22] This holds also for Fig. 6, where the word „cancer" is used, but anomaly in general should be understood.

5. A computer system that could emulate this will include one individual's ontological "stepping stones" into general knowledge about those phenomena. In order to make a characterization, it would shift into a design-based reasoning. This is not the literal pattern matching, but a way to automatically reason about *how processes manifest in form*, the slides of tissue. As with analogy, the results would come in the form of multiple suggestions, rather than definitive analysis – the final judgement comes back to the human expert, but this is support that can bring the right features to their attention, saves time and perhaps lives.
6. The interesting thing about thinking in terms of a 'design' is that it produces structure that sums up an entire process or situation. It embodies the higher-level reasoning currently absent in methodologically reductionist approaches (Gare, 2017).
7. In an analysis system, it would be a way to read an image and pick out evidence of anomalous and normal processes, not only based on the shapes, but also based on characterizations of overall histories of shape transformations.
8. The physician annotates what is happening in the slides as if by **assigning processes** *to the physical structures*, to tell future diagnosticians why they need to look for these physical features. He says: 'look for this physical effect, because it indicates the biology.' A machine capable of doing his job would also need to take into account some aspects of those processes. The biology teaches a machine (an 'expert') what to look for. **The biology and the physical structure is the *residue* of the process interactions** in the tissue and the resulting flow/structures produced.
9. The visual display is like a *weather map*, with causal imperatives displayed as a deduction of what is observed, because the cause is what is searched for. There is a causal lattice, where the cloud of shifting ontologies swarms around what is concrete and known. The dynamics of interest are in the way those clouds flow, rather than any particular cluster of nodes in them. A pathologist is looking for interactions and flows characteristic of malignant processes.

C. Question: imagine you are observing one of those Esher pictures with mazes of houses and staircases up and down which are all constructed in some logical, causal manner letting us to locally comprehend a context as normal, but from the global view, our stomach is turning back all over to tell us there is something abnormal in the whole picture. Imagine, you wish a computer to tell you the truth through such an image analysis. The question is *how to identify the switches between these local pairs feature-cause/process segments in the reasoning paths and the global one in the overall picture*. There are obviously *multiple 'perceptual switches'* telling us where are the absurd regions. We have not put noise in these pictures yet.

Based on this plan the physician structures the diagnostic findings and traces the treatment. S/he pays attention to specific details that support or contradict the therapy. For instance, there are 'stained' sections of tissue that the expert looks at to make a diagnosis of cancer (Fig. 3). The identification of such areas needs to be supported by the WLIMES diagnosis assistance system.

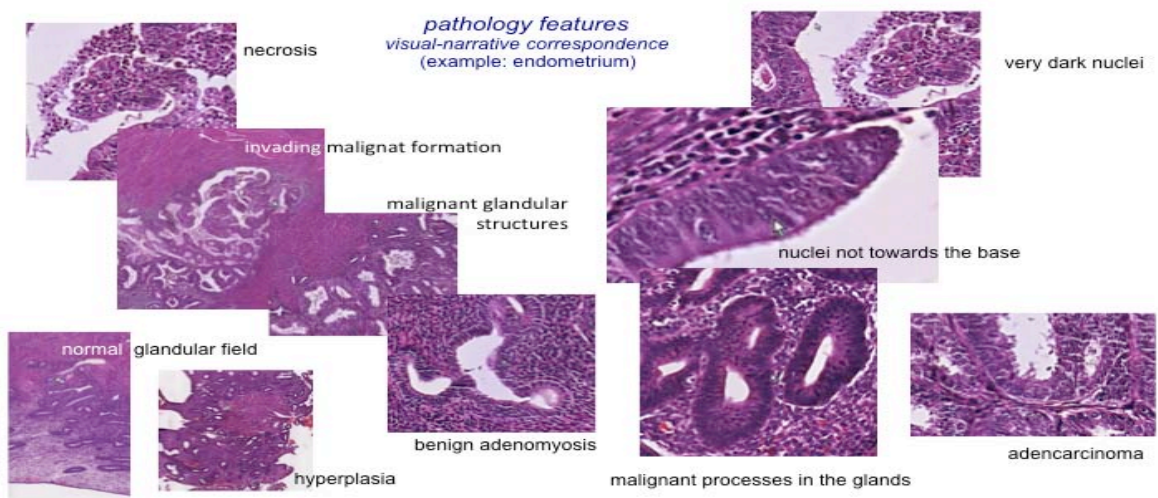

**Fig. 3**: *The Object: a visual-narrative correspondence in the WLIMES categorical ontology for WSI*

As the narrator goes through the image, it is clear that what s/he is looking for are the physical results of biological processes (the 'residue', the 'leftover', the 'spoors'). 'Lexical', textu(r)al are the objects with their structures and arrangement with others. Yet, the meaning of these constellations is of particular interest. Semantics is hard to discover in images of the world. It is the task of the scout who creates an entire process, a story based on the spoors left. The gaps in the story are filled by facts delivered by deduction and previous experience. The experts are actively involved in the creation of the WLIMES diagnosis and decision support loop (Fig. 4).

This process begins with the (machine) analysis of audio-visual records of Whole Slide Images which were a priori examined by experts. The idea behind is to identify a visual-narrative correspondence shared among a large number of experts. Herewith a hierarchical ontology map between the expert's narrative description of the artefacts in the images in terms of form, texture, context, position/situation etc. characteristics and the corresponding visual representations of these structures is going to be constructed and validated in multiple tries in order to extract and sort out the high-level dependencies between the observed anomalies (Fig. 5) and the underlying processes. This ontology map can be used to derive rules for an automated expert-like pre-diagnosis categorization of individual cases. Simultaneously, a relational database of "holistic" feature categorization will be prepared and constantly updated in terms of a phraseological dictionary for tissue structures, thus enabling the fine-tuning of the rules. The obtained results will be validated and updated in juxtaposition with the patient history knowledge base and other non-personalized data about "statistical average" from low-level feature extraction and classification systems (not present on Fig. 4). We call this entire process machine validation and decision support. The final result is screened by an expert pathologist before being stored in the individual patient's knowledge base. From there it can be later used for interactive case exploration, hypothesis generation and proof.

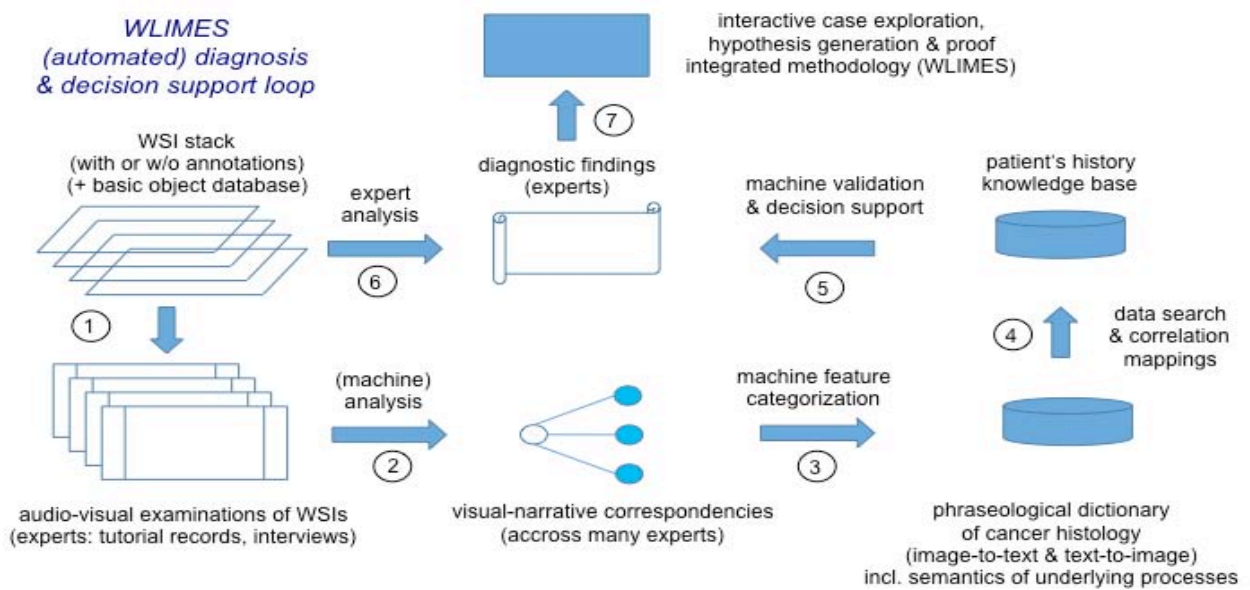

**Fig. 4**: *The Subject: a schematic representation of the WLIMES diagnosis and decision support loop*

But what is the nature of biology the machine needs to understand? Its reports of the images would look something like a meteorological satellite map. These are more interstitially depictions of thermodynamic urges, writing wetness, heat, wind, lamination, etc. A machine should read such an image and pick out evidence of anomalous and normal processes based on the shapes. There's always weather, and always events out of the norm. Cancer, for instance, is a particular conjunction of systems and conditions. In the general case, variation within normal samples is greater than between normal tissue and cancer. The indicators of cancer are very subtle and several ones of them need to be put together. We could also try to analyze not the image but its encoding, the language/description instead of the subject itself. The problem is however, that interpreting that 'code' needs concepts larger than physics. It is behavior at the word level, not at the concept level. What we need are the equivalent of fractals in the concepts space.

Art is different from design/architecture image because its transfer of information capability is not (only) related to 'aware information' but (also) to 'unaware information'. There is something the human brain is reading that the conscious mind does not. It relates to emotions and instincts that we cannot control directly. The comprehension of cancer images can have also this component, which could be related in some way to past memories and emotional flashbacks of the observer/expert. One way to solve this dichotomy is to teach a machine *to search for something 'interesting'* instead of searching for a specific thing. This implies not the use of an algorithm, which is a procedural concept, but some kind of **concept comprehension by the machine**.

Things may become very complex to reason about if we begin to consider additional details of women's physiology, which let all figures and structures vary in shape and density depending on the ovulation cycle, and the other parameters as age, heritage, psyche, etc. annotated in a personal profile. Concepts that can be used to reconstruct an expert's reasoning process when interpreting an image are those of:
- *higher cognitive and thought processes:* cognitive abilities like memory, spatial representation or higher cognitive processes up to consciousness and creativity (Ehresmann, 2012; Ehresmann and Gomez-Ramirez, 2015);
- *ideas which can be used to describe the pathologist's thought process*;
- the 'dynamic model' MENS (for Memory Evolutive Neural System) of a neuro-cognitive-mental system (Ehresmann and Vanbremeersch, 2009a/b).

(For the categorical treatment of these processes, cf. Appendix B.)

Therefore, our approach is simply to take an image and begin interpreting it. We wish to understand the meaning of the long-term trends of formation, how unseen influences produce visible 'residue', and what new developments can tell us about the past 'fossil record'. This analysis process thus includes also critical *implicit* factors. Here is where **multiple parallel reasoning threads can be validated against each other during the image (re-)construction, a sort of reverse engineering which imitates human retroactive reasoning. Which step to take forward and in which direction depends on what we know from previous experience – not just what has happened, but what we know *can* happen. This sense of what is normal and possible might change, particularly when we zoom to the level of personalized information. This could be managed by deep convolutional learning (in the machine case) and from its matching to the present situation in real time**. Yet, we should still keep in mind that the 'code', the 'weather map' we read and interpret is not the territory itself, only its visible products.

### 4.5. Perspectives

The WLIMES formalism is supposed to help us realizing a machine pre-diagnosis of whole slide images (WSI) that traces a histologist's reasoning process by learning from the analysis of previous records of audiovisual examination, textual reports, as well as first and third person patient narratives revealing the specific character of the underlying biological processes from the individual recognition and categorization of anomalies in the histological probes. The goal is to derive comprehensible and justifiable rules that can be used in the automatic WSI analysis and classification for medical diagnosis. To realize this we are going to develop techniques for connecting inductive learning methods with model-based, deductive and other approaches, thus enhancing and optimizing the conventional (deep) machine learning methods. These include in particular:
- *Use of expert knowledge*: procedures that introduce and integrate declarative or model-based a priori expert knowledge beyond the targeted selection of training data, such as, for example, initial and boundary constraints, as well as other methods;
- *Concatenation*: techniques for the sensible combination of the machine learning methods with other methods of artificial intelligence, as well as iterated complexifications processes leading to the emergence of more complex objects and links (a WLIMES approach to 'deep learning').
- *Explainability and transparency*: techniques for the user-friendly, transparent and comprehensible presentation of the solution path. Of paticular interest are appoaches, which increase the understanding of the functionality of the underlying processes and explain the path to the solution/diagnosis traceable and verifiable.

**Remark**: If instead of making a qualitative analysis of multiple expert interpretations of WSI, we try to use the WLIMES formalism for extracting quantifiable knowledge from static images in a bottom-up manner, we will realize that this technique is not suited for this purpose. A recent survey on anomaly detection via graphs (Akoglu et al., 2015) represents a synthesis of diverse methods, which unfortunately does not explain how graphs are used. When one tries to go into more details, s/he realizes that it is only about graphs in the restrictive sense of having only one edge (relation) between two adjacent vertices. This configuration is equivalent to matrices, which are suitable for performing explicit quantifications. The methods presented in the study do not extend to categories. The latter are able to address much richer relations/connections. They are difficult to disentangle in terms of matrices or relational databases, which can be later used to benchmark specific tissue formations. A quantitative analysis on a flat individual image also has its limits. The use of WLIMES for analyzing graphic images could help better in cases where instead of analyzing isolated WSI scans from biopsies of diverse patients, there is a sequence of scans of the same histological part of the same patient at different successive times. This is usually the case in radiology with applying CT/fMRI/PET image analyses on specific organs and parts of the body of the same patients. This would allow for a categorical trace of the evolution of the system (a tumor) over time. Static images alone as WSI samples are not well suited for a WLIMES analysis, unless we try to learn the expert's reasoning chain. Therefore, it is more suitable for applications in personalized medicine rather than for enhancing quantitative data in applications with statistical correlation. The strength of WLIMES is relational mathematics, not statistical analysis.

## 5. VLC4WLIMES: a visual language and calculus for WLIMES modelling and simulation

This section provides an overview of the VLC4WLIMES project, which is devoted to the design and implementation of a visual language and calculus for WLIMES modelling and simulation of multi-level living systems. It is the first step towards realizing WLIME(N)S shared models for personalized medicine discussed in detail in Appendix B along the implementation line outlined in Section 4.

### 5.1. Objectives

This project is going to develop and implement the core of the first WLIMES solution, its visual language and calculus (VLC) and an augmented reality interactive environment tool (Fig. 5), when approaching one of the crucial problems of signaling science: how the cell changes its categorical[23] "states" according to the general and specific dynamics of the incoming signaling pathways?

This problem impinges for instance in *tumor genesis and tumor dynamics*, in how the different tumor stem cells relate morphologically and biochemically among themselves and their environment – accompanying cloning cells of the tumor mass, lymphocytes of the immune system, fatty tissue, endothelial cells of the necessary new vasculature to develop, etc. Many individual cell's "decisions" have to be taken, implying thus, either accelerated growth, or retarded growth, or entering into the reproductive stage of the cell-cycle, or maintaining quiescence, or following senescence, apoptosis, necrosis, etc. All these global types of cell behavior and morphology are based on the information provided by multitude of signaling pathways that are continuously integrated and combined with the inner controlling information of the cell: metabolic, transcriptional, epigenetic, etc., (Marijuan et al., 2013, 2015), (Fig. 5). This problem of obtaining a general system view out from a multitude of elementary dynamics, each one with a number of variables and degrees of freedom and yet at a personalized level (s. note under Fig. 7), is currently "solved" by most system-biological approaches via mammoth systems of differential equations, often including a probabilistic component (e.g. Bayesian or Monte-Carlo methods) to compensate uncertainty – although the inner condition of continuity implied by the chemical Law of Mass Action does not hold in most pathways!

---

[23] The concept of "state" in this context is of graph-theoretical nature and should not be confounded with the more common definition of a cellular automaton. Let us take the specific mesh structure of an object, e.g. a multilayered organ tissue, or a concatenated process, e.g. an enzyme reaction. Let us represent it graph theoretically using nodes and connections between them where the nodes themselves are composed of larger structures — other graphs. Then this hierarchical nested patterning of nodes can hold a multi-relation structure which is pre-categorical in the sense that one has not specified composition laws for the edges. Then the scientist can allow some freedom of description introducing some composition laws of some of the edges.

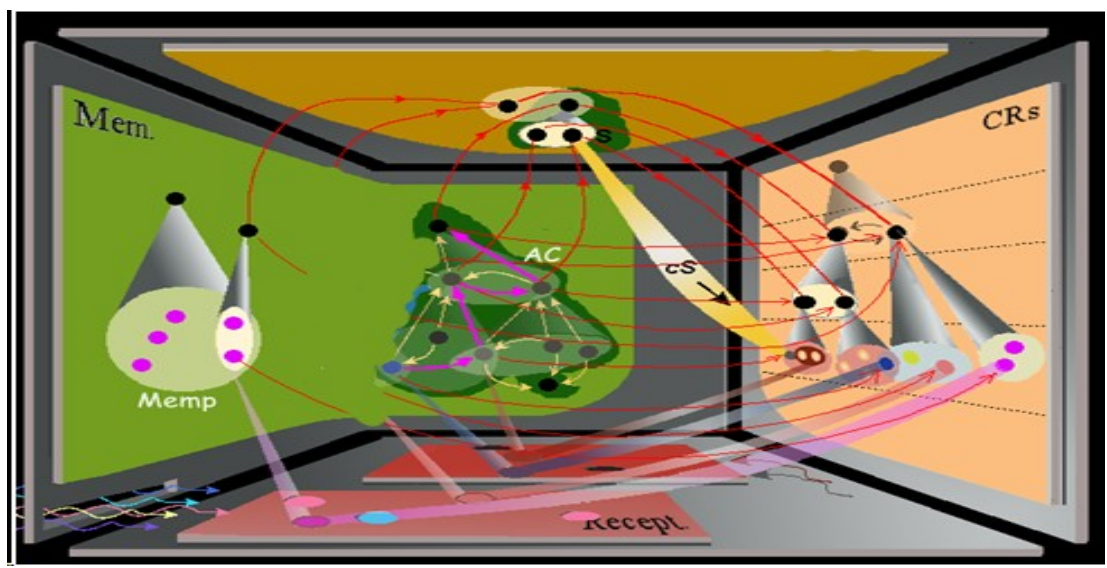

**Fig. 5**: *Modelling multiscale self-organization in a MES/WLIMES multidimensional space through dynamic categorical mappings: an evolving plastic heterarchical net of specialized subsystems (co-regulators, CR), each one of which has its own complexity, rhythm, logic and differential access to a long-term memory and develops by learning.*

Besides, many other procedural inconveniences are involved. In an attempt to sidestep this problem, new visualization tools (Premk and Bartz, 2007; Scharpe et al., 2008; Callaway, 2016) have been recently proposed to approach more efficiently fragments of the signaling landscape, such as the 'Minardo' chart of insulin action (Fig. 9), but the categorical problem remains untackled. The need for agile, expressive visualization of the dynamics of morphology, microbiology and signaling lies at hand.

The examples on figures 8-9 show how important is the continuous 'liveness' actualization of a signaling network that allows to draw conclusions, e.g. about the cellular intelligence (Ford, 2017). Therefore, the major goal pursued is the application of abstract biomathematical tools herein proposed to obtain an elegant cell-cycle state control based on vivid visualization and easy manipulation of objects and processes at multiple levels, integrating the most relevant signaling pathways, and covering all directions (bottom-up, top-down, inside-out and outside-in) of the information flows in the cells and the tissues.

By using WLIMES (Ehresmann and Simeonov, 2012; Simeonov and Ehresmann, 2017) enhanced with the mathematics of recursion and self-reference (Kauffman, 1987, 2015) it should be possible to assist physicians in delivering an overarching cancer diagnosis and therapy. To the extent that solving this core problem proves successful, it can be extrapolated to explore other anomalies, such as developmental and epigenetic disorders, immune system dynamics of microbial and viral infections, autoimmune diseases, etc. This kernel module can be re-used in other fields of complex system analysis and translational medicine to deliver more *systatic* (i.e. consolidating and probative, (Gebser, 2011; Simeonov, 2015)) models to understand truly causative relations and distinguish them from coincidental statistical correlations. The formal theory behind, WLIMES (Ehresmann and Simeonov, 2012), is the combination of both approaches, WLI (Simeonov, 2002b) and MES (Ehresmann and Vanbremeersch, 2007), which are complementary to each other. The mathematics is MES with the operational semantics provided by WLI.

The WLIMES theory uniquely addresses the *emergence, development, multiplicity* and *complexification* aspects of entire living systems. It is capable to capture natural phenomena as outright processes. But the necessary first step is the implementation of a visual language with syntax based on graph theory (of which category theory is an enrichment), and the diverse (hyper-)graph components operating in different categorical levels.

A project will develop a mathematical visual calculus with a graphical language syntax (interface) informed by an attributed computational semantics and demonstrated in clinical scenario of shared diagnosis a therapy in personalized medicine. In both the Wandering Logic Intelligence (WLI) the Memory Evolutive Systems (MES) theories temporality has been modeled as external linear clock time, which helps to measure and to compare the different specific temporalities of the CRs. This concept will be revised to incorporate *multiple phenomenological dimensions* of time (Cole, 1980; Strnad, 1980; Cole, 1981; Strnad, 1981; Matsuno, 1997, 1998; Varela, 1999; Matsuno and Salthe, 2002; Pegg, 2008; Price, 2012; Vargas et al., 2013; Matsuno, 2016; Vrobel, 2013, 2015), incl. anticipatory[24] ones, using adequate mathematical (Hoffman, 2013; Kauffman, 2015), cognitive (Atmanspacher and Dalenoort, 1994; Bateson, 2000; Canales, 2015) and computational techniques (Matsuno, 1995; Simeonov et al, 2013; Simeonov and Cottam, 2015). The challenge faced here is to overarch the complementary developments of both Goranson's Two-Sorted Logic (2SL, (Goranson and Cardier, 2013; Goranson et al., 2015; Cardier et al., 2017)) and Ehresmann's MES using a second order (subjective) logic (Jerzak, 2009; Khrennikov and Schumann, 2014; Simeonov, 2015). In fact, MES also uses three different types of logic for modelling biosystems like 2SL: i) the general logic describing the MES, ii) the specific logic of each co-regulator, and in higher cognitive systems, iii) the logic of the "macro-landscape" (or the "global landscape"), which accounts for agency, situations and phenomenology.

---

[24] The multi-temporal and multi-subject essence of bio-logic, e.g. in a predator-prey situation, should be also able to take into account the *anticipatory* (attentive or even *pre-cognized* and *meditative*) nature of life (Rosen, 1985, 1991, 1999) with respect to achieving specific goals. It is certainly worth exploring how a programmed direction of time from the future to the present is taking place. The roots of such a programmed (memorized) reverse direction of time as an observed/shared sequence of flexibly but willful scheduled and coordinated events can be observed from prehistoric tribe hunting rituals to present day project management. From incorporating these anticipation mechanisms into the multidimensional time concept, we should work to understand how *somatic models* are encoded, how they change over time, and what avenues exist for the scientists to reprogram the models that are a source of dysfunction. For example, an immune system dysfunction is often caused by misidentification of the self or other proteins as "pathogens". So, allergies, that can sometimes even be life-threatening, are entirely due to error in the models, not to the signaling, that activate the body's defenses in the event of an invasion. It's an incorrect prediction based on incorrect identification. If scientists could access those internal somatic models (or memories!), they could cure not only allergies but auto-immune syndromes and eliminate organ-transplant rejection. There are many other aspects of life and sources of dysfunction that are anticipatory in this way.

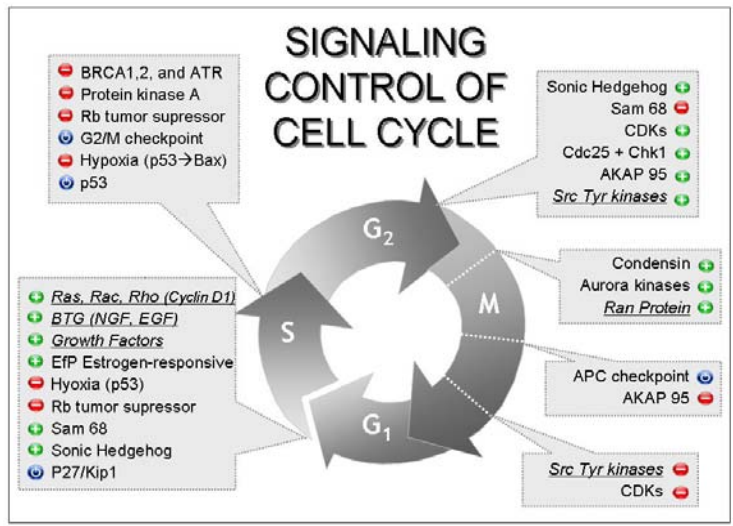

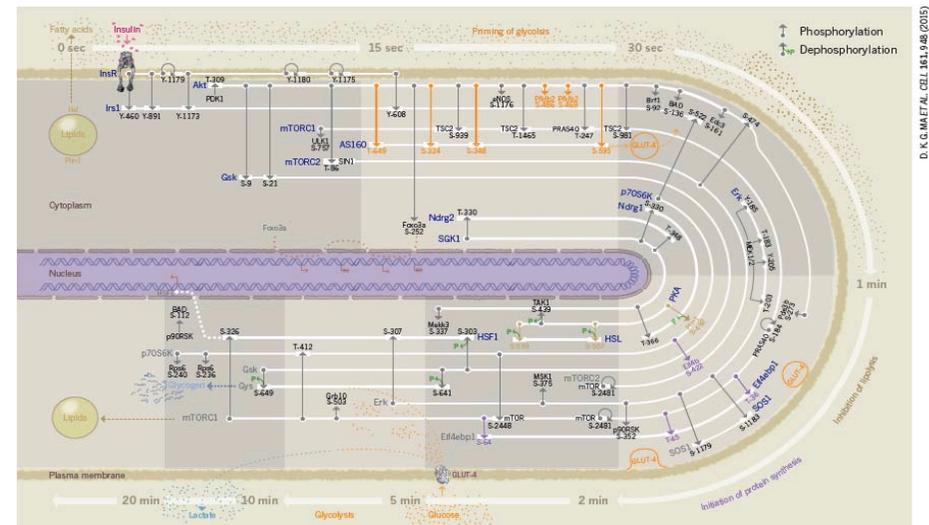

**Fig. 6:** *The global/all-embracing view: cell-cycle control (from (Marijuan et al. 2013), with permission).*

**Fig. 7**: *The local/specific view: a flat spatiotemporal action chart of insulinj/IGF signaling cascade of protein phosphorylation after a cell is treated with insulin* (from (Ma et al., 2015)).

Legend: A tight signaling control is established on the different phases of the cell cycle (*G*1: gap; *S*: synthesis; *G*2: interphase gap; *M*: mitosis) and on their respective transitions. The modular organization of the multicellular organism allows the space-time separation between cell-cycle phases, mediated by a number of controlling signaling pathways. The signaling control is ultimately based on a cloud of internal and external signals, usually of opposed signs (activators vs. inhibitors), that carefully regulate the reproductive and specialization trajectories of cells and tissues. In the figure, activating signaling pathways promoting progress of the cell-cycle bear the (+) sign, while the inhibiting ones bear the (-) sign. In the case of cellular checkpoints the sign is a 'switch button', as they can result in progress or in arrest, depending on the incoming factors. The signaling pathways associated to MAP kinases appear in italics.

Note: This figure illustrates the large (and continuously growing) number of parameters to be taken into consideration in conventional system-biologic approaches, which inevitably leads to combinatorial explosion. Even if we were able to identify all the system's components and track all their interactions we will end up with having as many degrees of freedom as states (observables), and the model will become a replica of the system itself and will cease to be a model. Therefore, category theory is needed to avoid this combinatorial explosion of coincidental correlations.

## 5.2. Related Work

The idea of having a visual modeling system for mathematics based on AI/ML, computer graphics and animation analogous to a sophisticated combination of the open source *Processing*[25] software sketchbook/language for coding in the visual arts and *Wolfram Mathematica*™[26] for technical calculations and plotting in the sciences is a powerful one. It can allow maximal illustration and exploration of abstract concepts and objects e.g. through haptic interaction. Of particular interest here is the visual representation of the formal semantics of mappings to compute emerging and evolving properties in multi-level systems/categories, which is the focus of the present project. Category theory (Mac Lane, 1998; Awodey, 2006) has been used as universal modelling tool to resolve complex problems not only in physics, engineering and design, but also in life sciences (Ehresmann and Vanbremeersch, 1987; Rosen, 1991, 1999; Ehresmann and Vanbremeersch, 2007; Louie, 2009, 2013; Ehresmann, 2017b). A number of higher-order logics, calculi and functional programming languages for categories have been also designed or adopted (e.g. Haskell, ML, Coq, OBJ, OCaml, CAM, Idris, Agda, as well as LISP/SCHEME/Clojure and Scala). *Globular* is perhaps the newest development in the field. The system[27] developed by Bar and Vicary at Oxford University (Bar and Vicary, 2016; Bar et al., 2016) is a very sophisticated tool[28] for visual modelling of higher category theory. Higher-order categories, today a well-studied domain, were initially introduced by Charles Ehresmann in 1963 and developed in several papers along with his wife Andrée until his death in 1979. Globular is tailored to visually represent complex computations in higher order (or 'multiple') categories such as 2- or 3-categories or monoidal categories which allow a simultaneous planar (2D) handling of several composition laws. Andrée C. Ehresmann has thought of introducing multiple categories in her formal theory MES, but she argues that they are not suitable for addressing the main questions, which WLIMES raises with using conventional categories for modelling living systems. Another approach of this work plan considers linking parts of this research to the one suggested by Goranson, known as Two-Sorted Logic (2SL), (Goranson and Cardier, 2013; Goranson et al., 2015; Cardier et al., 2017). Globular uses conventional programming algorithms and techniques to display interactions among composition laws of monoidal categories, while SBC – functional programming techniques for direct encoding of categorical interactions aiming their presentation as conventional deductions in the 2SL, i.e. with the converse intent. Therefore, n-categories and Globular can be adopted later with the necessity to address more complex problems in WLIMES, which requires the introduction of several compositions on the same set (for instance to compute interactions between different MES).

## 5.3. Work Plan and Methodology

This initiative is pioneering some new ideas in a couple of fields in computer science and mathematics. Its major advantage is the pursuit of a new realm interweaving mathematics with naturalistic computation for science. A research project is going to develop and implement a first prototype of a Visual Language and Calculus for WLIMES (Ehresmann and Simeonov, 2012; Simeonov and Ehresmann, 2017), VLC4WLIMES, as part of a novel mathematical formalism that underpins the construction of pertinent dynamic multi-level models for synthetic and natural biosystems. The proposed approach would help exploring the system (e.g. a patient or disease) as a whole analogous to a Prezi[29] graphics visualization, yet in a much more elaborated manner, using a Shared Augmented Reality Diagnosis Assistant (SARDA) environment which lets the researcher not only represent and view the system elements and their relations, but also adjust and test their interplay in connection with various hypotheses that will be then validated using external data.

---

[25] https://processing.org/
[26] https://www.wolfram.com/mathematica/
[27] http://globular.science
[28] https://johncarlosbaez.wordpress.com/2016/12/14/globular/
[29] www.prezi.com

The hope is that the outcome of this pursuit will enable the exploration of natural phenomena via interaction with(in) a new kind of synthetic dynamic flowchart environment being composed of (self-)evolving modules and their interactions. The construction of this tool will be realized in 2 stages. Each one of them is going to address the phases of a two-sorted iterative design procedure: i) specification/representation, and ii) validation/proof using WLIMES. As a first step, a diagrammatical visual language for WLIMES, VL-WLIMES will be defined and developed to facilitate bidirectional human-human and human-machine interactions, (Marriott and Meyer, 1998; Zhang, 2007).

Starting from a basic set of iconic symbols (objects) as pictorial representations of mathematical and computational concepts, this language will be developed to construct higher-level conceptual entities and relations of WLIMES distributed and shared in multiple dimensions (Fig. 5). They will be basically of graph-theoretical and category-theoretical nature, but will also include semiotic attribution (in addition to the syntax and semantics of the language grammar) for context-dependent processing that will vary depending on the application context. The next step will be the construction of the Visual Calculus, VC-WLIMES to operate on the language constructs (known as a compiler/interpreter in earlier 'static' language generations) following the WLIMES paradigm. The talk is of calculus and logic and not of compiler or interpreter, since some part of the logic/rules will be a priori defined in the visual language (graph) grammar used for modelling, but another will be extracted and learned (by the system) from the differences[30] between the model and the real data fed into the system to adjust that model and make it appear more realistic. Thus, the ontologies about the application domain should be (semi-)automatically derived and used to uncover and classify detected (recursive) patterns of objects/structures and processes in large-scale systems that cannot be usually extracted using human assistance. The envisioned functional operators to operate upon the structures of the emergent and extensible model in series of iterations will be also both pre-set and derived ones from the results of executing the basic WLI and MES principles upon the domain and language/calculus mappings/morphisms. They will define the foundations on which the new modeling methodology for designing complex biological systems, VLC4WLIMES, is going to be built on. The work can begin with investigating the use of well-developed open source platforms and tools for 2D/3D computer graphics design and animation used in advertisement and gaming, such as OpenGL[31]/WebGL[32], Blender[33] and Unity[34]. Another approach is to explore the suitability of commercial products such as those of the Audodesk[35] product line for engineering, architecture and cinematography (Maya™ and 3dsMax™).

Both options are targeting a visualization workbench for implementing model underpinnings for hypothesis generation and validation. This has been earlier applied in other fields of medicine and life sciences (Premk and Bartz, 2007; Scharpe and Lumsden, 2008). This time, however, the focus of using these graphical tools will be not on capturing the global *subject* of investigation in different stages, which is the case with 'Minardo' state charts, (Callaway, 2016), but on the applied *methodology,* i.e. diagnosis and therapy.

The proposed scope envisions the pursuit of the entire cycle of modelling and validation of complex dynamic systems, supported by the integration of real data, deep learning and high-performance computing, (Fig. 8). It is hoped that the fusion of computer-generated models with 'Big Data' and AI/ML methods might lead to more realistic, sophisticated and predictable models. The *key innovation,* which could allow realizing this goal, is the use of the same kind of 'object' virtualization and enhancement also for the conceptualization and validation of the methodology operating on these models, thus keeping aligned with the category-theoretical view on the diverse domains underlying this approach.

---

[30] and the "difference that makes the difference", Bateson (2000)
[31] https://www.opengl.org/
[32] https://developer.mozilla.org/en-US/docs/Web/API/WebGL_API
[33] https://www.blender.org/
[34] https://unity3d.com/
[35] http://www.autodesk.com/

The plan envisions the application of this methodology in life sciences and medicine, in particular for the visual representation and computation by mathematical transformations in complex dynamic systems. It could also be used in other domains of science and the humanities – where hierarchy-heterarchy, emergence and development are of interest (McCulloch, 1945; Merelli et al., 2012; Acotto and Andreatta, 2012).

Specifically, the following tasks need to be tackled on the way:

- to design and explore mathematical concepts from such fields as category theory, topology and quantum (categorical, temporal, modal) logic;
- to express dynamic (varying) hierarchical graph and category structures (that is, Evolutive Hierarchical Systems) with attributed properties capable to change the functions of the nodes and edges and convert patterns of connected objects into higher objects for modeling the emergence and development of complex natural and synthetic life systems, such as e.g. multilayer structures from systems and organs down to the cellular and molecular levels for studying biological phenomena and anomalies such as cancer and neurodegenerative diseases;
- to merge, compare, enhance and validate model generated morphological data with real ones from biopsy scans (Whole Slide Images, WSI) and radiology images (PET, fMRI, CT) in multiple iterations, thus teaching the resulting model to develop to a more realistic level, suitable for automatic diagnosis.

To achieve this goal, along with the basic technology research group of the envisioned project, it would be necessary to have a team of skilled programmers with substantial mathematical background able to master the scripting of the VLC4WLIMES into MEL, Python, OpenGL/WebGL, Scala and C++ for driving both the creation of meshes and their animation. Autodesk Maya™ 2017 provides appropriate tools in the MESH editor for parameter driven creation and animation. Combining MESH and MEL/Python scripting is expected to deliver the desired result of being able to construct nested VLC4WLIMES expressions in a multi-dimensional space that can be navigated through on monitors using standard PC peripherals and SARDA equipment. The latter should represent the initial state of the underlying system model. They will be then semi-automatically (re-)configured and processed through VLC4WLIMES to deliver a working simulation/animation of the underlying complex system that is going to be validated subsequently against a set of pre-cognized boundary conditions with having the opportunity to adjust them with real (measured) data from external sources.

In a first step the VLC4WLIMES implementation will explore the utilization of (interfaces to) standard system modelling languages such as SysML/UML and web/graphical APIs for rendering 2D/3D graphics WebGL/OpenGL, open source 3D content creation software such as Blender, as well as commercial tools and their APIs from the Autodesk product line involving the exploitation of programming languages such as MEL, Python, C++ and Scala.

Evidently, emergence and development, so typical for living systems, which are addressed within WLIMES cannot be captured a priori by these instruments. The realization of these properties will require the setting of individual approaches to implementation. Once VLC4WLIMES has been tested for a certain period of time in a certain domain it can be allocated to a specific dialect of VLC4WLIMES for that particular domain. In this way there should emerge a whole set of visual languages and calculi stemming from a basic VLC4WLIMES instance. The hope is that the outcome of this venture would be an augmented reality software tool to be integrated in various eHealth platforms such as the Charité Digital Medicine Platform (C-DMP), and more generally, as a prototype mobile AI solution of a Shared Augmented Reality Diagnosis Assistant (SARDA) for personalized medicine. It is supposed to offer a basic set of capabilities of the WLIMES theory that can be used to design and explore dynamic categorical models and simulations of the structure and behavior of complex multi-level whole-systems.

## 6. Outlook and Conclusions

As already said, *visual languages, computer graphics, animation, virtual and augmented reality* are well known and developed fields in gaming and cinematography. They have been often used in scientific visualization, in particular for representing mathematical concepts such as those in topology. This time the project should go one step further and use their capabilities to visually theorize, interact with and experience mathematical operations in an SARDA[36] scenario, thus realizing the WLIMES formalism in virtual oncology, cf. Fig. 8, left-hand side. The hope is that this innovation might lead to a new generation of analytical and modeling tools urgently needed to appease the dearth for creative and effective research automation not only in life sciences and medicine, but also in other domains.

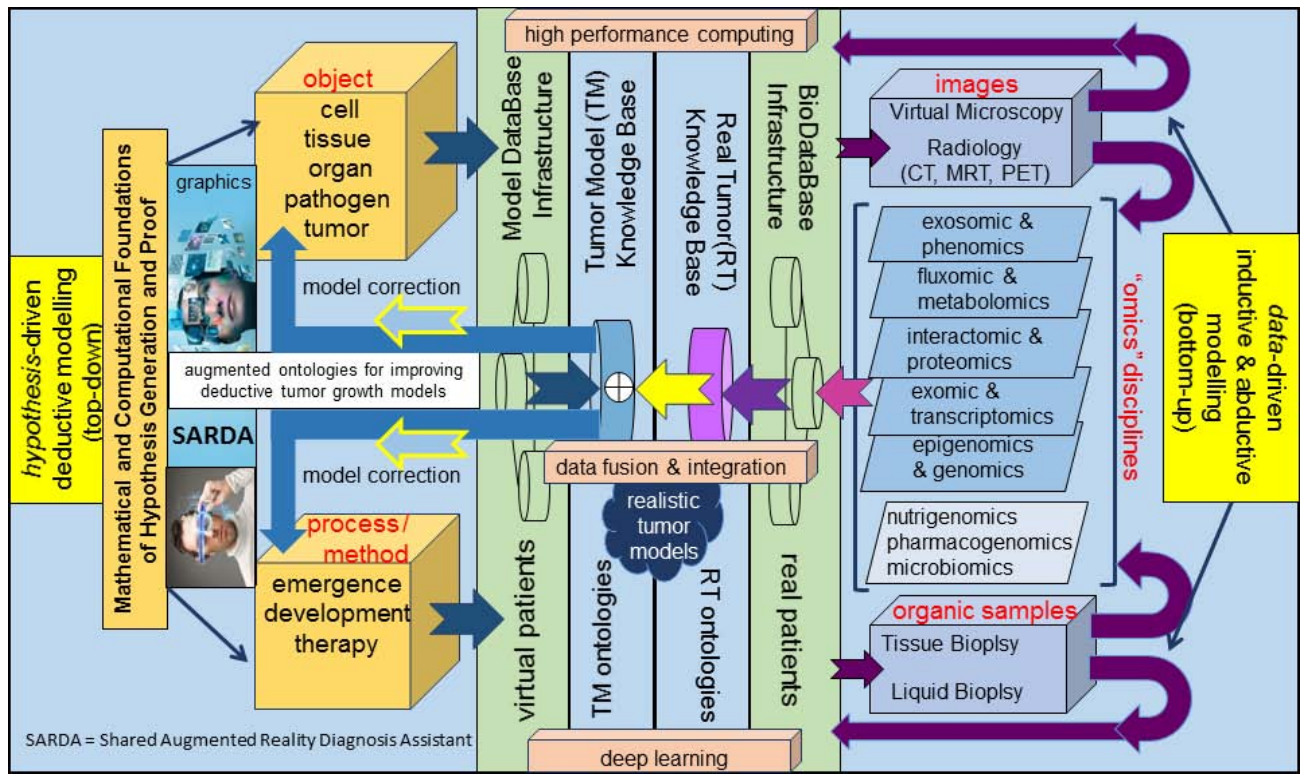

**Fig. 8:** *The potential contributions of Integral Biomathics and WLIMES to personalized medicine and virtual oncology (left-hand side)* through workflow integration of iterative object model and process simulation enhancement with real data:
i) 3D object reconstruction from samples and ii) interactive expert-machine methodology validation.

In particular, the WLIMES approach when enhanced with creative visualization and interaction with data has the potential to develop new hypothesis generation and proof tools in future personalized medicine (cf. Fig. 8, left-hand side). Its major advantage is its capability for smooth integration of mathematical deductive models into the traditional workflow of conventional inductive and abductive diagnostics and therapy.

In this process, the research plan envisions the iterative correction and enhancement of theory-driven models and simulations about complex natural phenomena with real data registered and generated by analytical tools, thus advancing such fields as virtual oncology, virtual virology, virtual immunology and many others.

---

[36] Augmented Reality Diagnosis Assistant

The WLIMES approach combines the advantages of a multi-scale multi-agents multi-temporality methodology (MES) based on a 'dynamic' category theory and a situation and context aware computational logic for active self-organizing networks (WLI) to systemically integrate theoretical and applied research. The results of this effort can be combined with other related research (Goranson and Cardier, 2013; Goranson et al., 2015; Cardier et al., ; Gunji et al, 2007, 2008, 2017; Marchal, 1991, 1994, 2000; Kauffman, 1987; Smoryński, 2002; Bolander, 2002; Perlis, D. 2006; Tozzi et al., 2017; Gunji, 2017) for use within the current methodology and practices of theoretical biology and personalized medicine to deepen and to enhance the understanding of life.

Characteristic for this approach is its innate relation as practical method, e.g. through *self-reference* or even *self-introspection,* (Kauffman, 2015), to both natural and artificial life systems. An earlier paper (Simeonov and Ehresmann, 2017) has reviewed WLIMES as modern research tool from the perspective of a set of archetypal concepts in ancient Eastern teachings. The 6 WLI principles and the various MES constructions and theorems have been used loosely as system design guidelines so far. Their integration within an "organic" scientific methodology for analysis and prediction of living system behaviors is a task for the future.

Key innovations to be addressed in future would be the inclusion of the formalization of: i) both the exo-physical (third-person) and endo-physical[37] (phenomenological, first-person) description of the system (Atmanspacher and Dalenoort, 1994; Ehresmann and Gomez-Ramirez, 2015); ii) a second-order dynamic logic for the observer-participant (Jerzak, 2009; Khrennikov and Schumann, 2014; Simeonov, 2015); iii) a multi-level system emergence and development (Kauffman, 1993; Kauffman, 2009; Merelli et al., 2012; Longo et al., 2012; Kauffman, 2015); iv) a multi-dimensional flow of time (Cole, 1980, 1981; Strnad, 1980, 1981) as well as v) treatment of anticipation (Rosen, 1985; Louie, 2010; Ehresmann, 2013, 2017abc) and vi) retrocausality (Pegg, 2008; Price, 2012; Matsuno, 2016). These characteristics are going to be investigated in a series of research projects towards creating a well-grounded general theory of living systems (Simeonov et al., 2012b) based on the WLIMES frame (Ehresmann and Simeonov, 2012; Simeonov and Ehresmann, 2017).

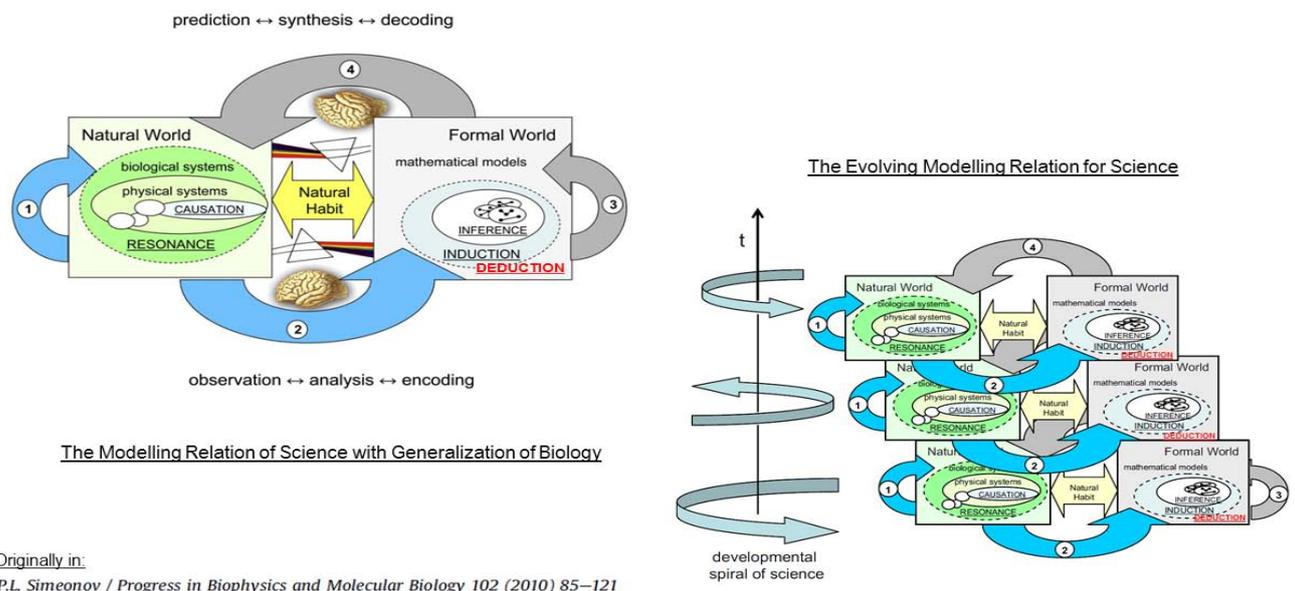

**Fig. 9:** *Giving priority to deductive methods in future life sciences and medicine.*

---

[37] formalized e.g. in the macro-landscape of the intentional CR (constructed using the archetypal core), cf. Section 4.1 and (Ehresmann and Gomez, 2015).

The inclusion of the deductive approach to life sciences (Fig. 9, marked in red color) is another novelty in the methodological concept of life sciences and medicine when compared to the initial scope of Integral Biomathics[38]. The goal is to develop this methodology, mainly through optimizing the iterative process cycle 1-4 and the self-referenced double arrow of natural habit finding its incarnation e.g. in the system architecture shown on Fig. 10. The work plan suggests that at the end of this venture there will be a prototype solution of a software SARDA tool with a basic set of capabilities of the WLIMES theory that can be used to design and explore dynamic categorical models (incl. phenomenological time) of the structure and behavior of complex multi-level whole-systems in life sciences and medicine.

This approach is expected to benefit future personalized medicine, particularly in its generic approach for consolidating top-down deductive theory-driven mathematical modelling with bottom-up inductive and abductive data-driven approaches, which is ultimately the major driving force behind Integral Biomathics.

The WLIMES theory allows us to model and trace not only of the *object* of investigation (Fig. 3, Fig. 8: upper left corner), but also the *subject* or process/method for investigation (Fig. 4., Fig. 8: down left corner). It enables the construction and the validation of the overall research process using the same mathematical abstraction mechanism allowing for adequate ontological representation and correspondence between both elements following the "Feynman principle" initially defined for quantum computation (Feynman, 1982), which is applicable to other research domains. When reasoning over multiple systems, states and times, a level of uncertainty emerges, but there are concrete top-down methods to handle this. These methods are the typical way of doing science, art/music, philosophy and even technology where reasoning about open world phenomena binds form to function and "(en)folds the act of observation into the model itself", (Cardier et al., 2017). Our approach thus stands in contrast to the often politically indoctrinated epistemological reductionism (Nagel, 1961; Gare, 2017) that has long been dominating science in the West. This is what we, in the West, can learn from Eastern scholars and other practices that emphasize systems thinking, jointly developing an extended framework of science unifying and reconciling the Eastern and Western traditions of thought, (Simeonov and Ehresmann, 2017; Cazalis, 2017).

*Acknowledgements*

*The authors wish to thank Beth Cardier, Ted Goranson, Nicolo Casas for the interesting discussions, which led to Section 4, as well as Manfred Dietel for his review and recommendations and Pedro C. Marijuán for the reference material and the critical remarks in Section 5.1. Our special thanks goes to Peter Hufnagl for the exciting discussion that led to the design of Fig. 8.*

*Author contributions statement*

This paper is the sequel of two papers jointly co-authored by Plamen. L. Simeonov and Andrée C. Ehresmann and Section 3 recalls some of their results. The new explicit contribution of Andrée C. Ehresmann for the present publication consists in providing Appendix B. The remaining parts of the main paper and Appendix A have been developed and written by Plamen L. Simeonov. They contain (in part) content from the postdoctoral thesis of Plamen L. Simeonov related to his qualification exam as a university lecturer (venia legendi).

---

[38] depicted in Fig. 3 and Fig. 4 on pp. 99-100 in (Simeonov, 2010)

**Appendix A: The WLIMES semantics as a wandering network emergence and maintenance via context and situation aware information processing and exchange**

This section provides an example for context-aware information processing. The terminology used is characteristic for mobile telecommunications. Further details can be found in (Simeonov, 2002b/c). The adopted WLI model for WLIMES uses shuttles (i.e. active packets / mobile code) to control the state in mobile nodes, netbots, semantic agents distributed in a multi-dimensional virtual model space. The following assumptions are guiding the WLI synchronization model:

1. Each netbot knows best its own configuration and state, as well as *how* and *when* to display it. This information can be encoded in shuttles and propagated throughout the virtual network. The netbot also maintains its own *reachability tree*[39] w. r. t. a particular transmission session, its role and each data flow transfer ending at it. A reachability tree can be dynamically verified and updated with the corresponding reachability trees (or parts of them) of other netbots. This "directed routing" information is periodically verified against and updated by the network topology patterns contained in a special kind of routing shuttles, *r-shuttles*, which periodically traverse the netbot.
2. Each netbot always provides *true*[40] (fair) information about its connectivity to other peers. Of course, some selection/filtering mechanisms might be applied to different subsets of nodes depending on some self-maintenance and performance optimisation criteria; it is not required that a netbot always tells the *entire* information about its state.
3. Netbots are supposed to be both communicative and *cooperative*[41], *but they can be also competing*. There might be different mechanisms to stimulate their cooperativeness or competitiveness in providing their resources to other netbots.

Let us now consider a scenario of the emergence of a wandering network. We point out that this example solely illustrates the capability of the WLI approach to handle routing, i.e. directed information exchange between the semantic agents in a comprehensive, straightforward manner. We assume that the routing state of this wandering network is completely described by the set of reachability trees (r-trees) $T_R$ of the individual netbots participating the network. The root of the r-tree is always the host netbot. The leaves and the intermediate nodes of the tree are the corresponding netbots from which the host netbot can be reached. Each netbot is responsible for:

1. maintaining its r-tree by collecting information from the shuttles traversing it;
2. forwarding shuttles to other destinations; and
3. reporting changes in the own r-tree structure, such as establishing new connections or cancelling old ones, to its neighbours.

In our model each node, except the root, represents an attributable virtual non-terminal element. This means that at every single moment the r-tree can be expanded or collapsed at such a virtual non-terminal. The synthesized attributes of this non-terminal represent the computed potential links to other netbots of the network.

---

[39] A reachability tree is the graph replacement for a routing table in telecommunications. In the WLIMES model it represents the surrounding context of or the situation. In general, netbots may execute several different tasks allocated to different subsets of the network. They are also serving as routers, i.e. shuttle directors, for other netbots. We used the term "reachability tree" instead of the well-known routing table for two reasons: (i) the ultimate goal of a routing algorithm is to determine the shortest path to each destination which may be the node itself; therefore, a routing tree rooted at the sink represents a loop-free set of paths which can be best maintained at the sink itself; (ii) routing trees contain *interconnection patterns* that can be easily verified against and matched with other patterns carried by shuttles.
[40] The treatment of un-trusted systems such as e.g. the Byzantine Generals problem is not part of this work.
[41] *Fairness rules* for netbot cooperation is an interesting research issue because of the limited performance of the mobile nodes which are required to serve as routers for their neighbours. In general, a trade-off between "routing for others" and the own tasks' maintenance should be considered. It is clear, that each node can serve only a limited number of neighbours depending on its own configuration and power consumption. In general, the more power the ship has, the more neighbours it can access and the more hops it can send its packets to.

Let us now go back to the construction of reachability trees in WLI and assume two single netbots, A and B, freely traversing the two-dimensional[42] model space. There is no connection established between them. Thus, each netbot contains only one single element in its reachability tree: itself, the root. Then, at some point of time both netbots approach each other within their access range. One of them, say A, initiates a connection protocol with the other netbot. The opposite side replies positively and the connection is established. Next, each netbot constructs a new branch of its reachability tree ending at the new neighbour.

At the next moment, a new netbot C approaches A and requests a connection. Upon a positive reply the connection is established and the local r-trees at each node are extended by the new branch.

However, this time both B and C are unaware of the fact that they may contact each other by letting A to route the shuttles between them. Therefore, netbot A is required to inform each one of its neighbours about the existence of other members[43] in the network. This is achieved by encoding and encapsulating the correspondingly "missing branch" information as executable *r-genes* (reachability genes) into the *r-shuttles*[44] (routing shuttles) which A transmits to its neighbours. Instead of a destination address and a TTL-counter (time-to-live), each r-shuttle is carrying an encoded tree branch called a *q-tree* (quest tree); it is required to traverse until being discarded at the end nodes[45]. The communication environment of the netbot can manipulate[46] both the q-genes and the r-genes of an r-shuttle in order to update their information, (Fig. 10).

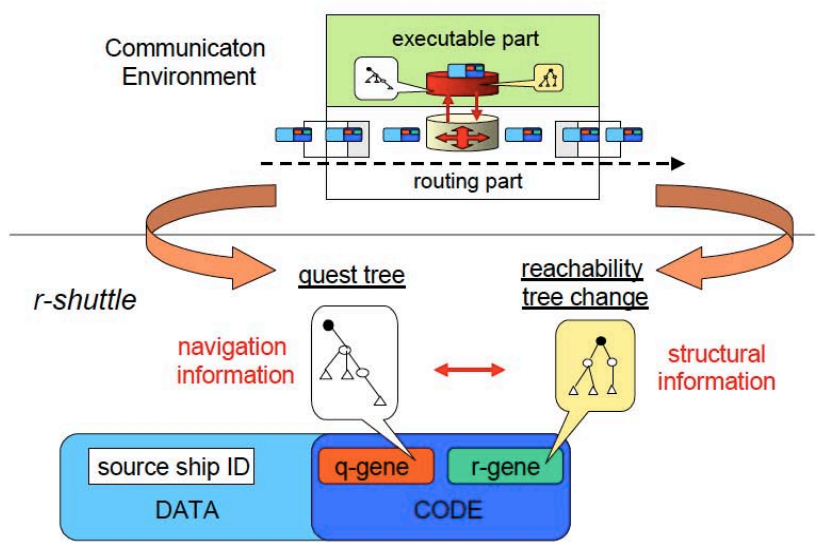

**Fig. 10**: *A communication environment manipulating the genetic structure of r-shuttles*

---

[42] In the general case we can have a multi-dimensional model space.
[43] We postulated in WLI that *fairness* and *cooperation* are a must. Yet, WLIMES allows for partly hiding and manipulating information for some reason.
[44] Reachablity shuttles in this scenario are much more likely to be regarded as dedicated semantic agents.
[45] Of course, an r-tree update method using an IP-like final destination and a TTL-counter could be also applied to guide the r-shuttles and limit their rotation in the network. However, since r-shuttles have a particular role to inform the netbots on their path about changes in the r-tree of their originating netbot, it does not make much sense to generate multiple instances of them at the originator for each single node in the branch (multiple destinations). Moreover, r-tree updates are supposed to occur much seldom than the actual packet transmission between the netbots. Furthermore, while traversing the branch, a part of it may simply disappear for some reason, so that the r-shuttle can be discarded at the intermediate node. In addition, we use this case study to demonstrate that the WLI shuttles may carry multiple instances of encoded structural information: here – one for the r-tree change of the originating node and one for the q-tree to be traversed by the r-shuttle. We claim that this executable information does not overload the r-shuttle because of the limited number of netbots communicating on the path.
[46] Please note again that fairness has the highest priority in autopoietic WLI networks, but not WLIMES netwoks.

In case that netbot C also has some neighbours it can route to, it is required to send this information via r-shuttles towards A, which in turn takes care to distribute it along the remaining branches of its reachability tree.

Generating a new branch of the r-tree on a netbot and dispatching r-shuttles to inform the neighbours about the change can be performed simultaneously. Besides, the same procedure is performed simultaneously on both sides of the newly established connection. These are two important advantages of the distributed WLI routing algorithm. We claim that by using the r-tree and q-tree types of encoding in the shuttles, the updated routing information is sent effectively to all affected nodes of the network. The computing overhead for encoding and decoding the r-trees should be considered as minimal because of the event related character of the reachability tree updates.

Figure 11 illustrates[47] the first two steps of the WLI routing algorithm[48], which we refer to as the *projection* phase:
  i)    Connect (X, Y) & Build (T'$_X$,T'$_Y$), and
  ii)   Inform (X, Y, T$_X$ ,T$_Y$).

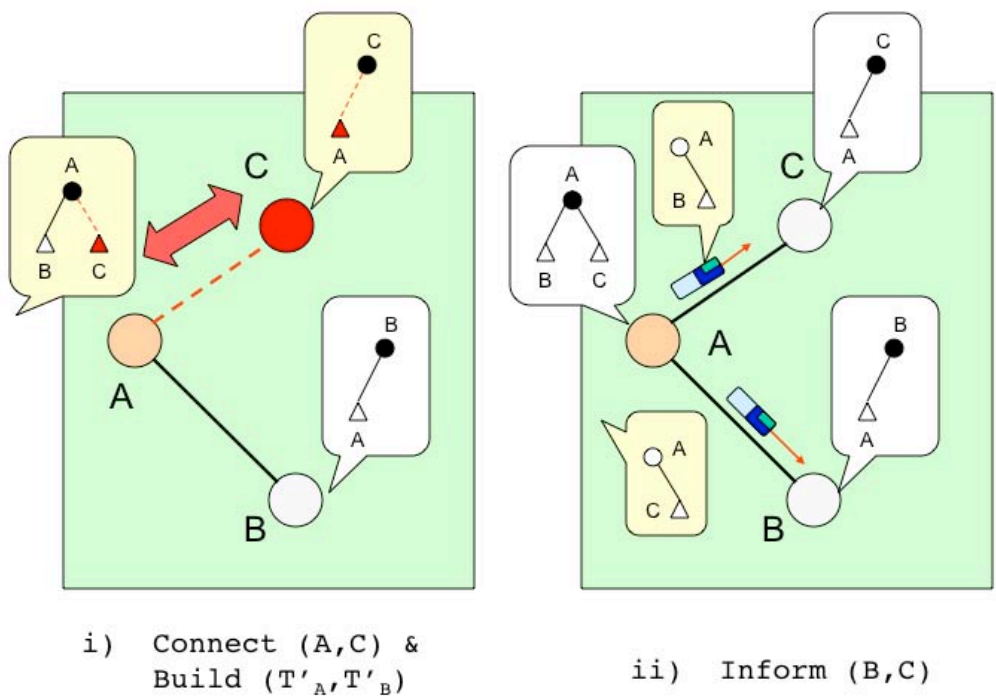

**Fig. 11**: *Projection*: building and transporting r-trees

We call the second phase of the WLI algorithm the *capturing phase*. It starts with the evaluation of the incoming shuttles and the expansion of the r-trees at the referred non-terminals by the "missed branches" encoded in the r-genes.

---

[47] The coloured circles denote *acting* nodes with the red one being the new netbot joining the network. The oval legends display the (parts of the) r-tree contents represented in the particular elements with the colour ones being active in the particular step of the algorithm. The shuttles on the figures are assumed to contain q-genes.
[48] The notion is taken for the general case of two netbots X and Y and their reachability trees T$_X$ and T$_Y$ . The prime sign upon T means the next state or the change of the reachability tree.

As soon as the r-shuttles arrive at their destinations, they are guided to the corresponding Communication Environment (CE) responsible for the link they come from. The CE then unpacks the "missing branch" information encoded in the *r-genes*, which are part of the executable code carried by the r-shuttle, and verifies it with the structure of its reachability tree. If the delivered information is redundant, i.e. the r-tree has been already constituted that way that the r-gene information represents a sub-branch of the netbot's r-tree (perhaps by a previously delivered shuttle from some other source), it is discarded. In case that the "new branch" information is a *really* new one, the CE takes care for expanding the r-tree at that virtual non-terminal which is assigned to be a root in the sub-tree encoded in the corresponding r-gene, (Fig. 12).

Finally, the reachability trees of all netbots are verified against each other (cf. Fig. 1, step iv) by broadcasting periodically r-shuttles containing the entire r-tree to the neighbours which analyse the incoming information on their side with the local tree structure and send back their feedback to the originating node. If no feedback is registered on a connection after some period has elapsed, the associated link is considered for failed and the change is reflected in the local r-tree and reported to the neighbours.

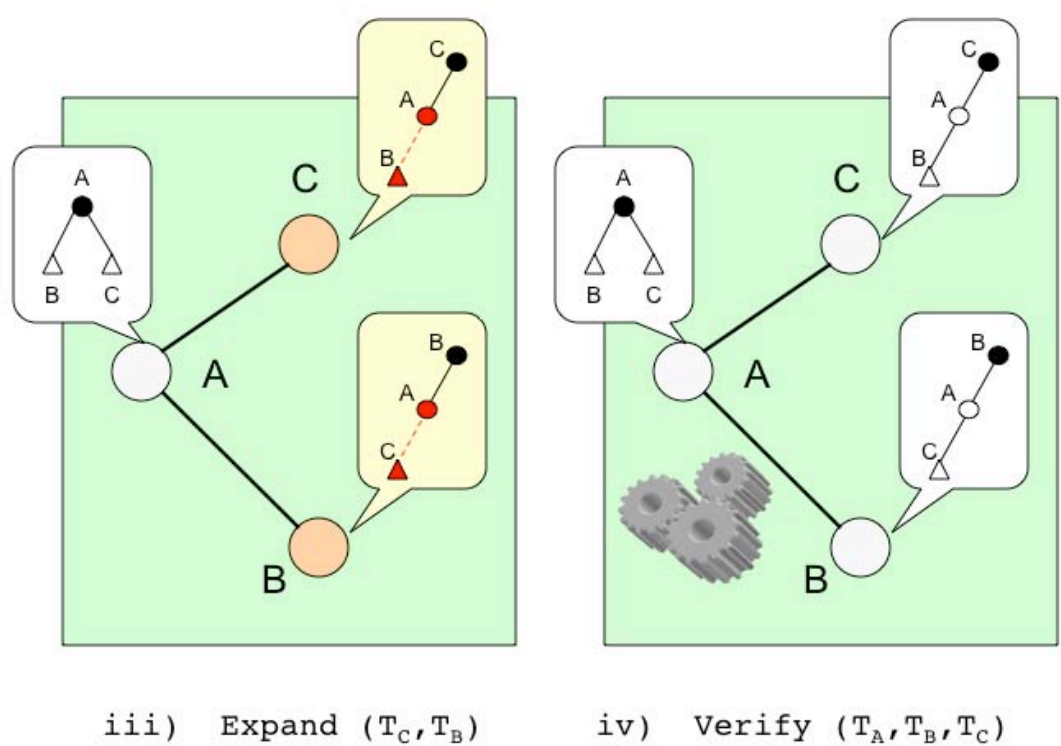

**Fig. 12**: *Capturing*: expanding and verifying r-trees

If the requested netbot is only an intermediate station on the path of the shuttle, the responsible CE updates the netbot's reachability tree by the r-shuttle's information and forwards it to the next hop on the shuttle's path. If there are any new structural changes on the shuttle's route ahead known by the CE, the shuttle's q-tree is updated[49].

---

[49] The r-tree can be also updated if there has been some recent changes in the source netbot connectivity known for some reason by the intermediate netbot, provided that the intermediate netbot is allowed to make such changes. For the moment, we assume the following: once being encoded in an r-shuttle, the r-tree remains unchanged (read-only).

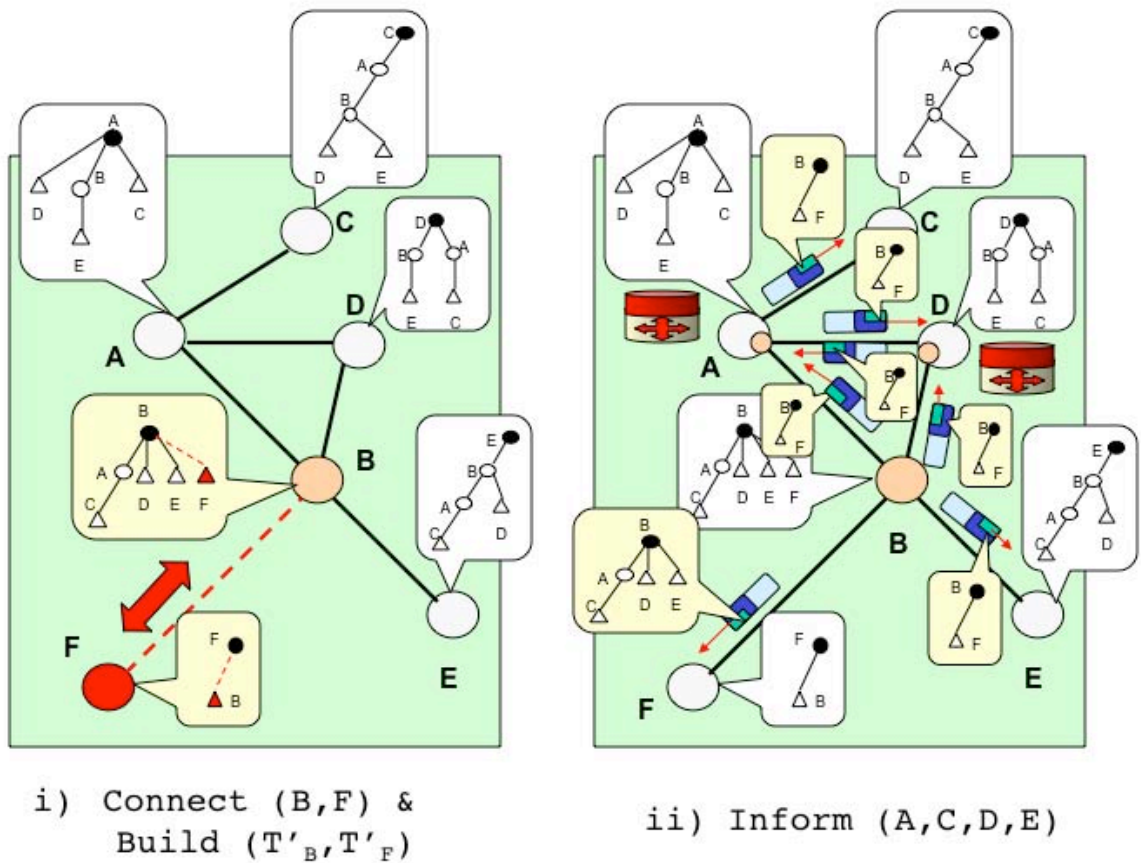

**Fig. 13**: *Projecting the inclusion of the 6$^{th}$ node of a wandering network*

Figures 13 and 14 illustrate the more complex example of updating the r-trees in a "three nodes ahead" evolving ad-hoc architecture. Note, that the same four steps – `Connect`, `Inform`, `Expand` and `Verify` – are taking place every time a new netbot joins the fleet. This is because of the distributed and parallel nature of the WLI algorithm with complexity[50] estimated to be **O(4 + m)**, where "m" is the maximum number of hops throughout all r-trees in all nodes participating the network.

---

[50] **Note**: the above assumption holds for flat hierarchies, i.e. the ones with only one level of interaction: the fleet or the local cluster as an ad-hoc distributed network (i.e. without a "head"). In case that new hierarchies are introduced, we have to adjust the algorithm to match the emerging levels and architectures of interaction. However, even then, the number of steps of the algorithm is progressive, yet limited which allows us to easily combine our approach with some well-known multi-level clustering strategies. *In each O(x) formula we imply **1 step for verification** at the end of the cycle.*

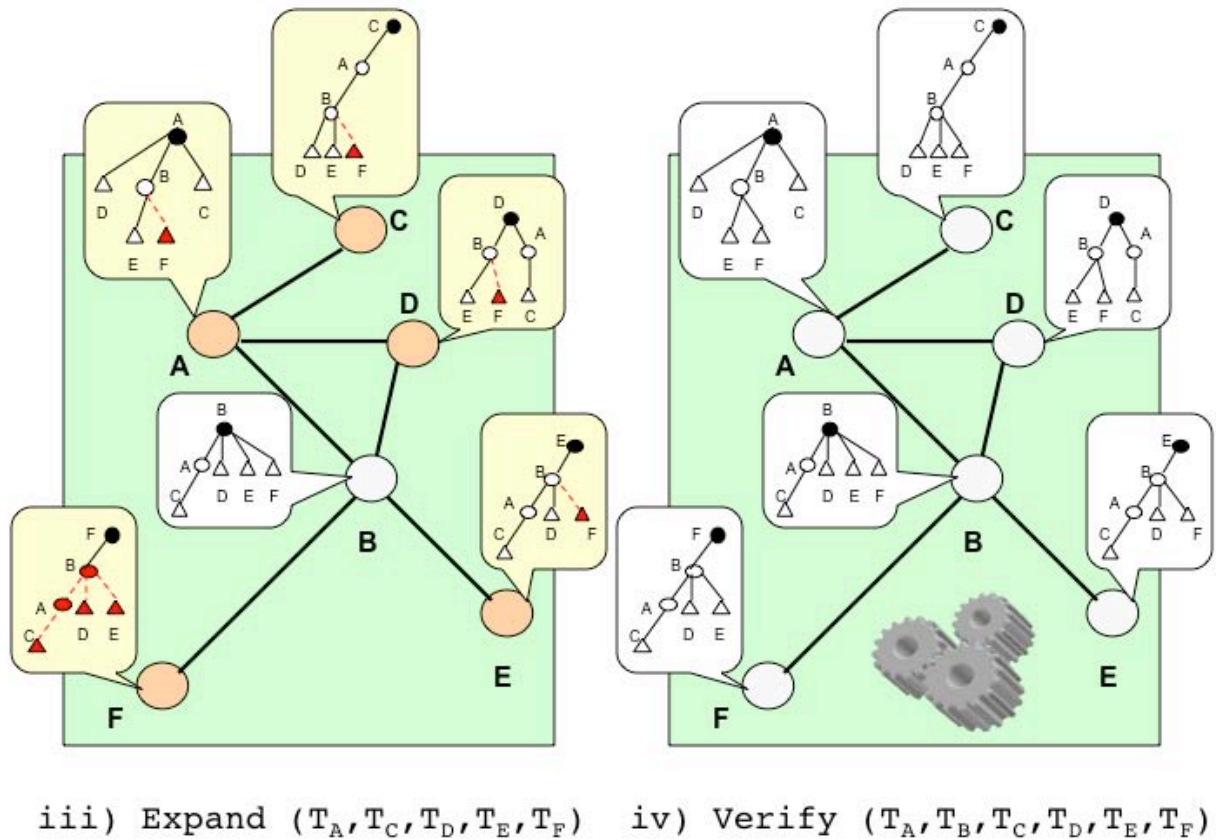

iii) Expand $(T_A, T_C, T_D, T_E, T_F)$    iv) Verify $(T_A, T_B, T_C, T_D, T_E, T_F)$

**Fig. 14**: *Capturing the inclusion of the 6th node of a wandering network*

Note that the small coloured circles inside the netbots A and D in step ii on Figure 15 denote the dual nature of the node in two consecutive steps: a) it acts as a *virtual non-terminal*[51], which expands its r-tree at some node upon evaluating the first incoming r-shuttle which delivers new information about the connectivity of that node, and b) it acts as a router/filter/replicator[52] for all other incoming shuttles with the same r-gene content from that node.

In case that a new, direct link is established between two netbots which already communicate[53] through other nodes, the "shortcut" is passed through as a new branch in the r-trees of the both netbots as shown on Fig. 15. Then, in the same step of the algorithm, each r-tree is depth-searched again to eliminate the dummy links and relocate the remaining branches on a shortcut path. For instance, in our case the link (B,D) is cancelled in the r-tree of netbot A, since there exists a shorter path from D to A. Analogously, the link (B, A) is cut through in the reachability tree of netbot D because D and A are now communicating directly, and not via B.

However, since A leads to C on that path and there is no other way for C to reach D, except through A, so the (A,C) branch is relocated, i.e. expanded, at the newly generated A. The complexity of this part of the WLI algorithm is **O(4).**

---

[51] The new link computed and assigned to the r-tree by the local CE after decoding and evaluating the r-gene of the incoming r-shuttle is *a new synthesized attribute* of the virtual non-terminal netbot.
[52] in case that meanwhile there has been established a *new link* ending at that node which is not considered in the q-tree of the incoming shuttle at the time of its encoding in the source nebot.
[53] i.e. both netbots are already present as virtual non-terminals in each others' r-trees.

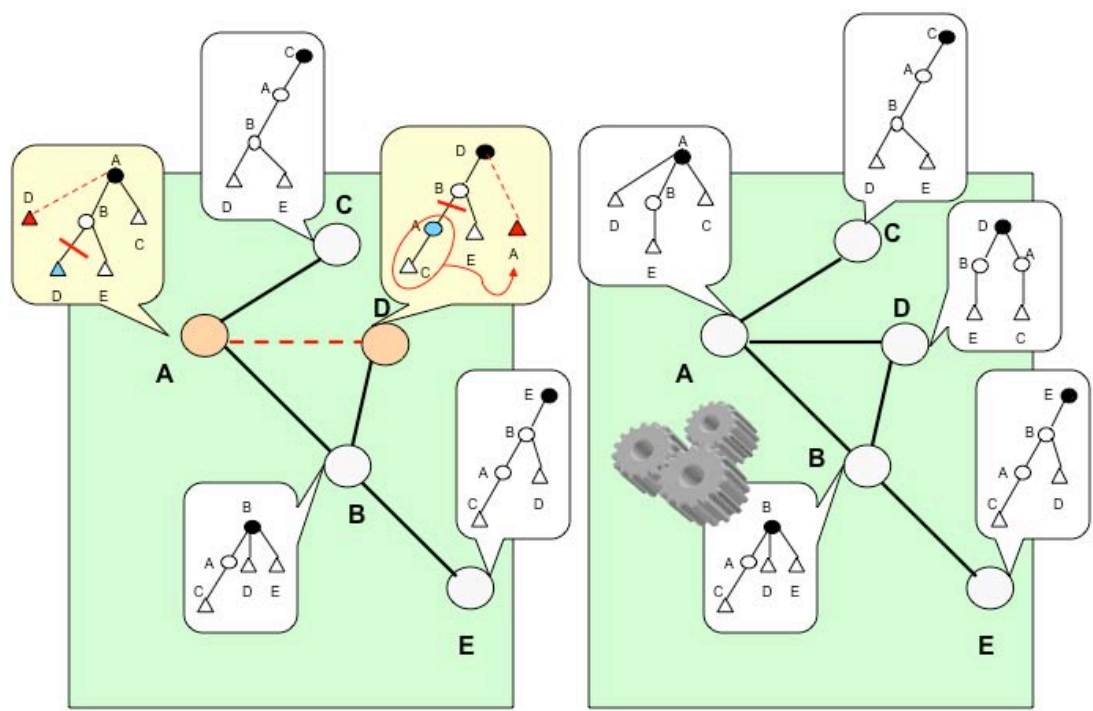

**Fig. 15**: *Introducing a short-cut*

Finally, let us consider the propagation of the r-tree changes when a netbot leaves the fleet for some reason (failure, movement, etc.) as shown on Figure 16. Firstly, the netbot can leave the network gracefully by informing its neighbours for his intention. Secondly, even if the netbot is going to leave the network spontaneously, this case can be reduced to the graceful one.

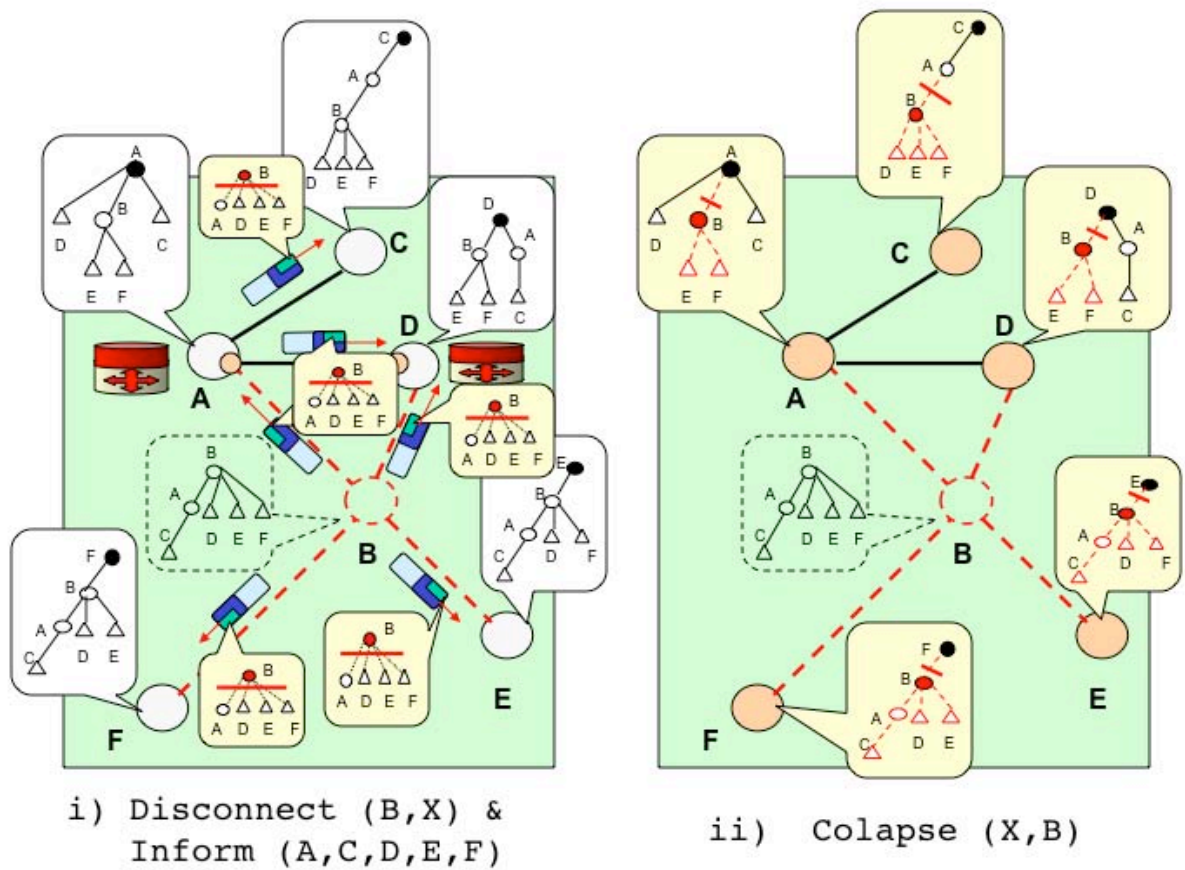

**Fig. 16**: *Projection of and capturing the exclusion of an intermediate node*

Each netbot can maintain an *alarm shuttle* (a-shuttle) containing an *a*-gene with the first level of the netbot's reachability tree, which includes the direct neighbours as leaves. The a-shuttle has a unique identifier that can be recognized by any netbot in the network. When the netbot intends to leave the network, it fires replicas of the alarm shuttle in all directions as a last action before going to inform its neighbours about this event. The a-shuttle is updated as soon as the first level netbot connectivity changes. This function is maintained in parallel with the rest of the netbot's activities and does not require a specific schedule[54].

As soon as the a-shuttle reaches a netbot, its a-gene is unpacked and the local r-tree is updated with the new information. If the same shuttle comes later, e.g. from another line, its content is simply discarded and the shuttle is forwarded to the outgoing lines. Alarm shuttles may implement a TTL data field such as in the common data packets to spare the q-gene and thus to limit their circulation in the network.

The complexity of this part of the algorithm is **O(3+m)**, where "m" is the maximum number of hops throughout all r-trees in all nodes participating the network. This value includes the firing of the shuttles (1), the collapse of the r-trees in each netbot (2), and the verification of the r-trees (3), which may take several steps depending on the newly emerged topology of the network, but though regarded by us as a single linear step.

---

[54] For instance, it can be performed each time a new link is established or an old one is cancelled.

**Appendix B: Written in collaboration with Jean-Paul Vanbremeersch.
Application of MENS, D-MES and WLIMES to personalized medicine[55]**

An important problem in our modern society is to develop a personalized medicine paradigm, which while being at the center of the health system, considers the "singular relation' between:
 (a) the patient with his/her general physical condition and disease symptoms, but also psychological state, affects and reactions toward preventing or fighting illness (e.g. accepting or not a consultation, a treatment, etc.); and
 (b) the physician s/he consults who will develop his own (situation) *landscape* (as we are going to explain later), possibly asking for complementary examinations or even hospitalization of the patient.

This actor couple functions well in everyday practice. However, in difficult cases (e.g. cancer, and other non-communicable diseases like obesity and diabetes, cardiovascular and chronic immune and kidney deficiency, etc.), the diagnosis/treatment process necessitates a series of interactive exchange levels between different agent groups that can be addressed to provide a *shared* compositional evaluation of the patient case (not restricted to imaging decision help as in usual diagnosis help systems) but also the family and various support groups to comfort the patient. Thus we will operate in a WLIMES, which should include a sub-MES representing this involved community. In this MES each actor develops his/her own landscape ("roadmap" of the medical case), and all of them together create a (shared) macro-landscape where the situation can be analyzed while accounting for all its (physiological and psychological) aspects. This is possible since the original MES methodology proposes tools for describing higher cognitive processes such as memory, thought, decision, action and anticipation (Ehresmann, 2017abc), at the individual (MENS model: Ehresmann and Vanbremeersh, 2007; 2009), phenomenological (Ehresmann and Gomez-Ramirez, 2015) or collective level (D-MES model, Béjean and Ehresmann, 2015).

**B.1. A brief overview of the MENS model and its Archetypal Core**

The Memory Evolutive Neural System (MENS) is a Memory Evolutive System (MES), accounting for the functioning of the human neural and cognitive system at different (micro, meso, macro) levels of description and across various timescales. It describes how different brain areas interact to generate flexible 'mental objects' (such as concepts and procedures) and cognitive processes of increasing complexity, up to learning, consciousness, thought, anticipation and creativity. MENS takes advantage of a common process in brain dynamics that is the formation and preservation of more or less complex and distributed neural (hyper-)assemblies, whose 'synchronous' activation is associated to specific mental processes. This association is not 1-to-1 due to the "*degeneracy*[56] property of the neural code", emphasized by Edelman (1989[57], p. 50), which is modelled in MES by the *Multiplicity Principle*. The mental representation of a stimulus is the common 'binding' (modelled in MENS by a colimit) of the more or less different neural patterns, which it can synchronously activate either simultaneously or at different times.

---

[55] The main ideas of this Appendix are developed in a paper by A. Ehresmann and J.-P. Vanbremeersch (to appear).
[56] Indeed the term "degeneracy" is a bit unluckily chosen by Edelman in his papers and explained as *"degeneracy, the ability of elements that are structurally different to perform the same function or yield the same output"* (Edelman and Gally, 2001).
[57] MENS has initially be developed as a mathematical model/translation of Edelman's book, which had a major influence on the work of Ehresmann and Vanbremeersh (2001, 2007, 2009). Edelman's Theory of Neuronal Group Selection (TNGS) perceptual categorization is directly related to the organisation of memory and learning. According to it memories are realized as reentrant strengthening of synaptic connections in the brain. The latter can come to the level of long-term potentiation (LTP) by reinforcement (learning) during the re-telling the story that would create what we call permanent memories. This is how modern AI and Machine Learning techniques touch the mathematics of the Memory Evolutive Systems through Edelman's TNGS.

Mathematically, a neural system can be modelled as an Evolutive System NEUR, (Ehresmann and Vanbremeersh, 2009a/b), (Fig. 17). Its components (also called *neurons*, if there is no confusion possible) model the physical neurons as nodes of a graph with the links being the synaptic paths between them. MENS is a MES having NEUR as its level 0 and obtained from it through successive complexification processes. This means that a component C of level n+1, called *cat(egory)-neuron*, is a multi-faceted component which is the colimit in MENS of each (hyper-)assembly of (cat-)neurons at levels less than or equal to n *synchronously activated* by a given mental object. Such a cat-neuron has several ramifications down to the level 0, which constitute its multiple "physical realizabilities" (Kim, 1998). We say that C is *'activated'* at a time *t* if the neural base of one of its ramifications is physically activated in the neural system; then this activation diffuses to increasing levels up to C by assembling the corresponding ramification; this operation requires some delay, called the *activation delay* of C, which increases with the level of C. Cat-neurons of higher levels represent increasingly complex mental objects and processes. In MENS the links between cat-neurons of level > 0 are both simple and complex links, while the links between (cat-)neurons of level 0 model synaptic paths.

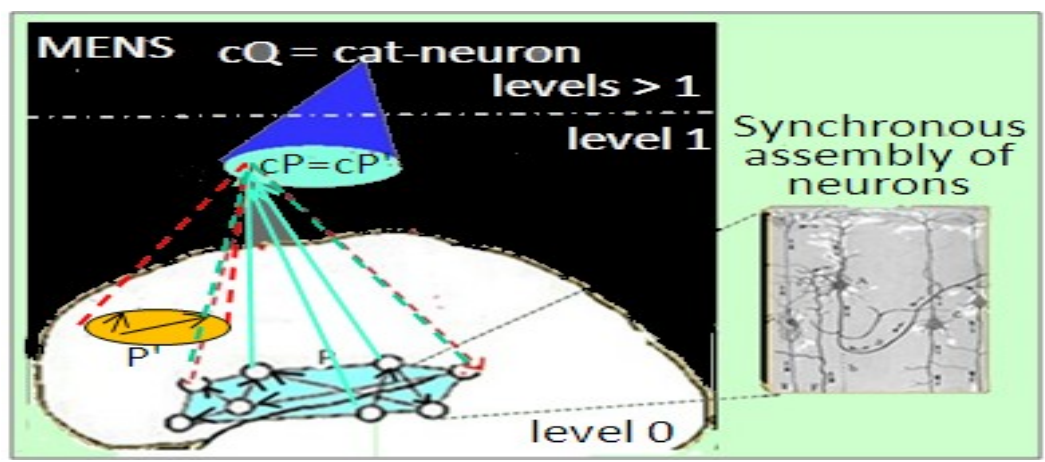

**Fig. 17:** *The hierarchical model MENS which has the Neural Evolutive System, NEUR, as its level 0 and is obtained from it by iterated complexifications.*

### B.1.1. The Memory and its Archetypal Core

As in any MES, the dynamic is modulated by the interactions between co-regulators, which, in MENS, are based on brain modules of different sizes and complexity. These co-regulators participate to the formation of a hierarchical memory Mem. An item (e.g. image) *I* activates patterns of cat-neurons in different co-regulators CR (e.g. related to shape, size, colour, density…); it leads (via the Emergence Theorem) to the formation of a colimit of the pattern in the landscape of each CR, called CR-r*ecord* of *I*; and the pattern consisting of the different CR-records binds in MENS into a colimit cat-neuron called the *record of* I. This record will be recalled (through the activation of one of its physical realizations) if I is later presented. Among the records we also discern records modelling different kinds of *procedures,* which can activate a pattern of effectors. In Mem we distinguish different sub-systems representing respectively a *procedural memory,* a *semantic memory,* and a higher integrative level called the *Archetypal Core.*

(i) The s*emantic memory* Sem gradually develops through the classification of records into invariance classes with respect to some of their attributes modelled by co-regulators CR (e.g.: items activating the same CR-trace on a CR-color, or a CR-shape). An invariance class is represented by a cat-neuron, called *concept* (e.g: 'color red') modelled by the projective limit in MENS of the CR-trace of each instance of the class. For more details, please refer to (Ehresmann and Vanbremeersh, 2007; 2009a, 2009b).

(ii) The human neural system contains, in the cortex, a topologically central *Structural Core* (SC), represented in (Hagmann et al., 2008), which has strong connector hubs, and, which plays an important role in functional integration; in NEUR it is modeled by an evolutive sub-system SC. In MENS, this SC allows constructing (via iterated complexifications) a higher level subsystem of the memory, called the *Archetypal Core* (AC), which is able to integrate and intertwine records and concepts of different modalities (sensorial, proprioceptive, motor, emotional, procedural…). Archetypal records and concepts are multi-faceted cat-neurons, with ramifications having their (neural) base in SC, which memorize recurrent experiences and realize *deep learning*. They are connected by loops of strong and fast complex links, which self-maintain their activity for a long time.

### B.1.2. Formation of a macro-landscape

An increase S in attention (e.g. analyzing an image) activates the neural base in SC of some archetypal cat-neurons A, whence the later activation of such A. This activation triggers, through archetypal loops, a self-maintained activation of a large pattern of AC, that propagates to lower levels through first a decomposition P of A, then, via a switch, to another decomposition Q of A, and down to lower levels through unfolding ramifications. Whence comes the activation of a large domain in MENS, which persists for a certain period. Such activation is transmitted through links to *intentional co-regulators* $CR_i$ (which are co-regulators based on associative brain areas and directly linked to AC). These $CR_i$ then cooperate and, with their links, act as a macro-Co-regulator to construct a *Macro-Landscape* ML (Figure 18). This ML is an Evolutive system, which connects and extends the landscapes of the $CR_i$, and lasts longer due to the self-maintained activation induced by AC. It develops through the processes: (a) propagation of the activation through loops in AC, unfolding of ramifications and switches between them; (b) recall of past or/and lower level 'events' by unfolding ramifications; (c) sharing of information by the $CR_i$'s.

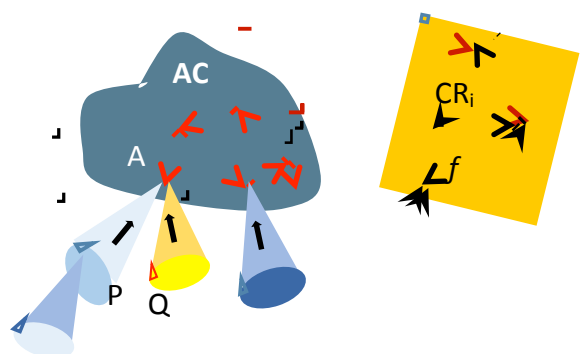

**Fig. 18**: *The Archetypal Core AC allowing the formation of a Macro Landscape (ML), which has for components the curved lines.*

Thus ML acts as a 'mental space' in which is operated (by abduction) a *retrospection process* for analyzing and 'making sense' of the present situation and its possible causes. It reflects and integrates processes of different temporality and complexity; some are conscious, others are 'non-conscious' (e.g. instinctive motor or perceptive behaviours, emotions and affects, reflexes), still others relate to *phenomenal data.* Then a *prospection* process can be developed in ML, still using the motor role of AC, to select an adequate strategy by constructing virtual landscapes in which strategies can be tried with evaluation of their risk of dysfunction. For more details, please cf. (Ehresmann and Gomez-Ramirez, 2015; Ehresmann, 2017c).

We are going to show how these processes will be at the basis of the decisions of the different actors interacting in a medical diagnosis and treatment process.

## B.2. Building the macro-landscape of a physician concerning a specific patient

During his/her medical studies a physician develops a large body of medical knowledge, which s/he continues to increase and amend in order to remain informed during his/her professional life. In particular, different diseases and pathological patterns are memorized as archetypal concepts, which have various ramifications, corresponding to different classes of symptoms, which a given patient might present.

Let us suppose that a woman visits a physician because she has some health problems. How will the physician arrive to a diagnosis and a treatment proposition? S/he will act according the following plan:

1. *First impressions.* As soon as the woman enters the practice, the physician's attention is drawn to her specific appearance, behavior, speech, etc. which are documented in sufficient detail: e.g. her general attitude, if she moves dynamically or seems depressed, how is the color and structure her skin, her hair, etc. If the physician (say, a man) already knows the woman, he can remark some changes with respect to the previous records or memories. In the MENS of the physician, the attention activates the structural core SC, including various (lower or intentional) co-regulators, and these aspects are inscribed in their landscapes.
2. *Oral and clinical examination.* Then the physician asks the woman how she feels, what are the symptoms for which she comes. This is the patient's own report which is also documented. Finally, the investigation is completed by a clinical examination (e.g. via ultrasound), which reveals other aspects of the disease such as the physical appearance of a lesion or an organ (its texture and form, the physical measurements of some anomalies, etc.) and the patient's response to touching; such characteristics of the disease cannot be easily recognized by quantifiable means and necessitate the involvement of a physician's practical experience.
3. In the physician's MENS, all the aspects collected in steps 1 and 2 are bounded together (formation of colimits), thus activating one or several archetypal records A through the up-folding of ramifications. By diffusion of the activation, a macro-landscape (ML) is formed, allowing for a retrospection process, which also takes into account information about similar former situations stored in the Memory of the physician. Let us note that the first more or less vague impressions of the physician are re-activated in this ML and can take an important signification in relation with the new discovered facts. What is called the 'intuition' of the physician results from this re-surging in ML of brief evanescent aspects in the landscape of some co-regulators, which re-appear in ML and are gradually made more precise by confrontation with later facts.
4. *Diagnosis and treatment.* If the different indications and information point to a specific pathology A, then the prospection process in ML searches for a known treatment. If several treatments (with more or less secondary effects) are possible, the physician will discuss with the patient to select which one she prefers. This is done in the frame of a larger macro-landscape uniting their two macro-landscapes (cf. Section 4.4.). If there are several possibilities (or none), the macro-landscape is extended and the retrospection analysis must be supplemented by asking for external lab examinations (e.g. biopsy, radiology, scans, etc.).
5. *Role of laboratory results.* Later, the physician will receive the results of the lab examinations. If s/he is not trained to interpret them, s/he may consult a specialist (pathologist, radiologist, etc.) and accepts his/her findings and conclusions, unless there is some reason that the diagnosis is not well compatible with other facts he has detected in the clinical examination; in this case further examinations can be prescribed. Finally, the physician structures the diagnostic findings and convokes the patient to discuss them with her, proposing an adequate treatment if possible or referring her to a specialist or to a hospital if necessary.

Section 4 above discusses a case study of shared image analysis following a similar procedure.

### B.3. Development of collective diagnosis and therapy decisions

We have said above that, while many medical problems can be solved by the singular dialogue patient-physician in more serious cases the intervention of other more or less specialized actors is required to reach a well-adapted collective decision on the case and to support the patient. Now let us consider a WLIMES containing all these actors. As shown in (Béjean and Ehresmann, 2015) and in (Ehresmann, 2017), MES provide the tools to study collective decision/action processes by analyzing how a more or less heterogeneous group G of people (here the different actors in a specific medical case) can arrive to an agreement in spite of their different backgrounds and initial opinions. Let us consider a group G composed of (some of) the above-mentioned actors. They build a team to collectively discuss the situation of a particular patient, attain a common/shared diagnosis and investigate and evaluate possible therapies. At the beginning of the meeting, the participants have different individual expertise and knowledge gained by direct experience or learning, conscious or latent opinions about the disease and/or the patient. The procedures that each member of the team would favor are not always compatible. The problem is to reach a compromise acceptable and shared by all and beneficial for the patient.

In the MES modeling of the social system, each member M of the group G can be considered as a co-regulator by itself, while the subsystem containing the members of G and the links representing the communication between them will be considered as a macro-co-regulator MG. About the patient, the memory of the MES should contain specific data not only about the present pathology but also about the medical history and even about the physical, mental and psychological personality of the patient. To arrive at a compromise, the idea is, by analogy, with the constructions made in MENS:
     (i) to define the notion of an Archetypal Pattern AG shared by the group G (as proposed in Ehresmann, 2017) with properties similar to those of the Archetypal Core of an individual; and
     (ii) by using it, to extend the landscape of MG (still called *macro-landscape*) through the sharing of information and knowledge that make sense of the situation, and to search for adequate solutions up to reaching collective decisions about the best therapy.

### B.3.1. Development of a G-archetypal pattern

The members of the G team exchange information and learn to share significant knowledge (explicit or tacit) related to the patient's situation, the possible diagnostics/therapies, their values, and possibly even some affects. Initially, a *multifaceted record A* such as a polysemous concept can have different meanings for two members of G; by exchange between them, they can share their views and give to A a common enriched meaning encompassing both so that A becomes a G-*shared record.* (This notion is stronger than that of a "boundary object" as defined by (Star and Griesemer, 1989).)

In the MES modeling the social system, the meaning of A for a member M of G depends on the ramification(s) through which M can recall A. Thus, A becomes a G-*shared record* if different members M of G can recall it through the unfolding of the same ramifications. The G-*Archetypal Pattern* of the group G is defined as an evolutive subsystem AG of the memory of the MES having for components G-shared multifaceted records of higher-complexity order, combining patterns of significant knowledge of various modalities (explicit, tacit, or latent). These G-archetypal records are connected by strong complex links which form archetypal loops self-maintaining their activity for some time. (The development of AG is a consequence of the Emergence Theorem.)

**B.3.2. Analysis of the situation in the macro-landscape**

The G-archetypal pattern AG plays the same role for the whole group G as the archetypal core AC for an individual[58] and it has the same properties; in particular, the activation of part of AG diffuses to a large part of the system, and this allows (as in the case of MENS in Section 4.2.2.) for the progressive extension of the *macro-landscape* ML which presents a common "mental space" for the group G. In it, AG acts as a motor for the development of collective decision processes by sharing not only explicit knowledge but also some tacit knowledge (e.g., tacit procedures, skills). Translating the constructions made in MENS to this case, we show that the macro-landscape ML contains the landscapes of the M's, interconnects them and extends them both spatially and temporally. In it, current observations and recent events can be related to past more or less similar cases, allowing to sense and making sense of the present situation, its trends and its possible evolution. This 'retrospective' analysis phase is followed by a prospection process for collectively establishing the diagnostics, searching for possible therapies, 'virtually' evaluating their expected results and finally selecting one of them. The group deliberations should be precisely consigned in a report, including the different options that have been discussed (even if not retained), as well as the final diagnosis/ therapy used and their later results. Here we have mainly considered the case of a group G of actors discussing the medical situation and trying to find jointly the best solution. The process would be analogous if the aim of the group is to inform and psychologically help the patient and his/her family.

**B.3.3. Development of a WLIMES assistance system[59] for personalized medicine**

The collective decision process for diagnosis, therapy and patient's support can be improved by increasing the amount of data that the advisory group(s) can share. For developing a personalized medicine decision support system in a medical institution (hospital, clinic, nursing-home, etc.), we propose the creation of a large database W containing the records of the WLIMES assistance system, for each treated patient featuring:

(i) Detailed information on the patient of any kind (medical, biological, mental, psychological, including personality and habits), kept up to date;

(ii) A report of the treating physician for each health problem of the patient about the symptoms, their progression, results of eventual complementary examinations, and if necessary, a collective decision about the case, incl. the reports of the involved groups.

Naturally such information should be kept confidential. Another well-protected database D is constituted from the first one (W) after patient data anonymization and classification of the major patient and disease characteristics. This database can be allocated within or interconnected with other already existing hospital automation systems. Now, when a new difficult case emerges, its details will be entered into D and (by automatic search, similar to the way histological slides are sorted), compared with the other cases in D, which demonstrate a larger number of similar characteristics (e.g. 75%). These cases will be taken apart and given to the groups examining the case, so that they obtain more information to consider and benefit from the results obtained by former groups, learning from their experience (also recorded with a final evaluation). Finally, the decisions and results of the new cases will also be transcribed in the database D. In this way the database D will evolve, making the WLIMES collective decision process more efficient, since errors made before are going to be corrected.

---

[58] The activation of part of AG will simultaneously provoke activations in the archetypal cores of a member M of G; however, these activations are of different natures, since they specifically depend on the personality of each M.
[59] The idea of a WLIMES decision support system based on an evolving database was inspired by the "prevention of risks" algorithms introduced by Dr. Jean-Paul Vanbremeersch to evaluate the residents of a nursing-home for elderly dependent persons (EHPAD in French, personal communication).

**Appendix C: Glossary of terms**

| | |
|---|---|
| 2SL | Two-Sorted Logic |
| AC | Archetypal Core |
| AG | G-Archetypal Pattern of a group G of people |
| AI | Artificial Intelligence |
| CA | Causal Agent |
| C-DMP | Charité Digital Medicine Platform |
| C-DPP | Charité Digital Pathology Platform |
| D-MES | Memory Evolutive System for Design |
| CR | CoRegulator |
| CT | Category Theory |
| CT | Computer Tomography |
| CU | Control Unit |
| DNA | DeoxyRibonucleic Acid |
| fMRI | functional Magnetic Resonance Imaging |
| IB | Integral Biomathics |
| INBIOSA | Integral BIOmathics Support Action |
| Mem | Memory in a Memory Evolutive System |
| MENS | Memory Evolutive Neural System |
| MES | Memory Evolutive Systems |
| ML | Machine Learning (in AI) |
| ML | Macro Landscape (in MES/MENS) |
| MP | Multiplicity Principle |
| netbot | (or "ship"): dynamically reconfigurable (physical) network element in a wandering network |
| PET | positron emission tomography |
| RCI | Resource Control Interface |
| RP | Resource Platform |
| SARDA | Shared Augmented Reality Diagnosis Assistant |
| SC | Structural Core |
| SCF | Service Control Function |
| SCI | Service Control Interface |
| SDF | Service Data Function |
| Sem | Semantic Memory |
| shuttle | active information packet of data and executable code transmitted between netbots in a wandering network the form of n-genes (or "network genes"); the latter are capable to pack and transmit an entire network infrastructure along with its functionality in terms of a virtual system model from one physical location to another. |
| SM | Switch Matrix |
| SN | Service Node |
| SNC | Service Node Controller |
| VLC | Visual Language and Calculus |
| VLC4WLIMES | Visual Language and Calculus for WLIMES |
| WLI | Wandering Logic Intelligence |
| WLIMES | Wandering Logic Intelligence Memory Evolutive Systems |
| WLIME(N)S | Wandering Logic Intelligence Memory Evolutive (Neural) Systems |
| WN | Wandering Network |
| WSI | Whole Slide Image |
| VC | Visual Calculus |
| VC-WLIMES | Visual Calculus for WLIMES |
| VL | Visual Language |
| VL-WLIMES | Visual Language for WLIMES |

page 52 of 56

page 53 of 56

page 54 of 56Simeonov, P. L., Ehresmann, A. C., Smith, L. S., Gomez-Ramirez, J., Repa, V. 2011. A New Biology: A Modern Perspective on the Challenge of Closing the Gap between the Islands of Knowledge. In: Cezon, M., Wolfsthal, Y. (Eds.) *ServiceWave 2010 Workshops (EDBPM 2010)*, Ghent, Belgium, December 2010. *LNCS 6569*. 188-195. Springer-Verlag, Heidelberg. http://link.springer.com/chapter/10.1007%2F978-3-642-22760-8_21. http://www.cs.stir.ac.uk/~lss/recentpapers/Simeonovetal2011.pdf.

Simeonov, P. L. 2010. Integral Biomathics: A Post-Newtonian View into the Logos of Bio. *Progress in Biophysics and Molecular Biology*. 102(2/3): 85-121. Elsevier, ISSN: 0079-6107. DOI: 10.1016/j.pbiomolbio.2010.01.005.

Simeonov, P. L. 2002a. The Viator Approach: About Four Principles of Autopoietic Growth On the Way to Hyperactive Network Architectures*,* Proc. FTPDS'02 | IPDPS'02, April 15-19, 2002, Ft. Lauderdale, FL, USA, IEEE Computer Society, 320 - 327, ISBN:0-7695-1573-8. DOI: 10.1109/IPDPS.2002.1016528. http://ieeexplore.ieee.org/iel5/7926/21854/01016528.pdf.

Simeonov, P. L. 2002b. *The Wandering Logic Intelligence: A Hyperactive Approach to Network Evolution and Its Application to Adaptive Mobile Multimedia Communications*. Ph.D. Dissertation*,* Technische Universität Ilmenau, Die Deutsche Bibliothek. http://d-nb.info/974936766/34.

Simeonov, P. L. 2002c. WARAAN: A Higher-Order Adaptive Routing Algorithm for Wireless Multimedia in Wandering Networks, 5th IEEE International Symposium on Wireless Personal Multimedia Communications (WPMC'2002), October 27-30, 2002, Honolulu, Hawaii, USA, 1385-1389, DOI:10.1109/ WPMC.2002.1088407. http://ieeexplore.ieee.org/iel5/8154/23649/01088407.pdf.

Simeonov, P. L. 1999. EP0957645 **–** *Network Element in an Intelligent Telecommunications Network.* European Patent.

Soto, A., Longo, G., Miquel, P.-A., Montevil, M., Mossio, M., Perret, N., Pocheville, A., Sonnenschein, C. 2016. Toward a theory of organisms: Three founding principles in search of a useful integration. *Progress in Biophysics and Molecular Biology. Special Issue "Principles for a Theory of Organisms"*. 122(1) 77-82.

Smoryński, C. 2002. Modal Logic and Self-Reference. In: *Handbook of Philosophical Logic,* Vol. 11. Kluver Academic Publishers. ISBN: 978-90-481-6554-4. 1-53.

Star, S. L., Griesemer, J. R. 1989. Institutional Ecology, 'Translations' and Boundary Objects: Amateurs and Professionals in Berkeley's Museum of Vertebrate Zoology, 1907-39. *Social Studies of Science*. **19** (3): 387–420. doi:10.1177/030631289019003001. http://innovation.ucdavis.edu/people/publications/Star%20Griesemer%201989%20SSS-19.3-387-420.pdf.

Strnad, J. 1980. On multidimensional time. J. Phys. A: Math. Gen. Vol. 13 (1980): L389-L391. http://iopscience.iop.org.

Strnad, J. 1981. Once more on multi-dimensional time. J. Phys. A: Math. Gen. Vol. 14 (1981): L433-L435. http://iopscience.iop.org.

Torday, J. 2017. Resolving Ambiguity as the Basis for Life. *Journal Progress in Biophysics and Molecular Biology. Special issue on Integral Biomathics: The Necessary Conjunction of the Western and Eastern Thought Traditions for Exploring the Nature of Mind and Life.* ISSN: 00796107. Vol. 131C. Elsevier. (in print)
54